\documentclass[12pt]{article}
\usepackage{epsf,amsfonts,amssymb,epsfig,amsmath}
\renewcommand{\text}[1]{#1}

\newcommand{\be}{\begin{equation}}
\newcommand{\ee}{\end{equation}}
\newcommand{\ben}{\begin{displaymath}}
\newcommand{\een}{\end{displaymath}}
\newcommand{\bea}{\begin{eqnarray}}
\newcommand{\eea}{\end{eqnarray}}
\newcommand{\bean}{\begin{eqnarray*}}
\newcommand{\eean}{\end{eqnarray*}}
\newcommand{\nn}{\nonumber \\}
\newcommand{\ba}{\begin{array}}
\newcommand{\ea}{\end{array}}
\newcommand{\bi}{\begin{itemize}}
\newcommand{\ei}{\end{itemize}}

\def\l{\lambda}
\def\a{\alpha}

\def\e{\epsilon}
\def\s{\sigma}
\def\e{\epsilon}

\def\w{\wedge}
\def\r{\rho}

\newcommand{\dd}{\mathrm{d}}

\begin{document}

\makeatletter
\renewcommand{\theequation}{\thesection.\arabic{equation}}
\@addtoreset{equation}{section} \makeatother

\baselineskip 18pt

\begin{titlepage}

\vfill

\begin{flushright}
Imperial/TP/2007/OC/03\\
\end{flushright}

\vfill

\begin{center}
   \baselineskip=16pt
   {\Large\bf Spacetime singularity resolution by M-theory fivebranes:
   calibrated geometry, Anti-de Sitter solutions and special holonomy metrics}
   \vskip 2cm
      Ois\'{\i}n A. P. Mac Conamhna
   \vskip .6cm
      \begin{small}
      \textit{Theoretical Physics Group, Blackett Laboratory, \\
        Imperial College London, London SW7 2AZ, U.K.}
       \end{small}\\*[.6cm]
      \begin{small}
      \textit{The Institute for Mathematical Sciences, \\
        Imperial College London, London SW7 2PE, U.K.}
        
        \end{small}
   \end{center}

\vfill

\begin{center}
\textbf{Abstract}
\end{center}

\begin{quote}
The supergravity description of various configurations of
supersymmetric M-fivebranes
wrapped on calibrated cycles of special holonomy manifolds is
studied. The description is provided by solutions of
eleven-dimensional supergravity which interpolate smoothly between a
special holonomy manifold and an event horizon with Anti-de Sitter
geometry. For known examples of Anti-de Sitter solutions, the associated special
holonomy metric is derived. One explicit Anti-de Sitter solution of M-theory
is so treated for fivebranes wrapping each of the following cycles:
K\"{a}hler cycles in Calabi-Yau two-, three- and four-folds; special
lagrangian cycles in three- and four-folds; associative three- and
co-associative four-cycles in $G_2$ manifolds; complex lagrangian
four-cycles in $Sp(2)$ manifolds; and Cayley four-cycles in $Spin(7)$
manifolds. In each case, the associated special holonomy metric is
singular, and is a hyperbolic analogue of a known metric. The
analogous known metrics are
respectively: Eguchi-Hanson, the resolved conifold and the four-fold
resolved conifold; the deformed conifold, and the Stenzel four-fold
metric; the Bryant-Salamon-Gibbons-Page-Pope $G_2$ metrics on an
$\mathbb{R}^4$ bundle over $S^3$, and an $\mathbb{R}^3$ bundle over
$S^4$ or $\mathbb{CP}^2$; the Calabi hyper-K\"{a}hler metric on $T^*\mathbb{CP}^2$; and
the Bryant-Salamon-Gibbons-Page-Pope $Spin(7)$ metric on an
$\mathbb{R}^4$ bundle over $S^4$. By the AdS/CFT correspondence, a
conformal field theory is associated to each of the new singular
special holonomy metrics, and defines the quantum gravitational physics of the
resolution of their singularities.

\end{quote}

\vfill

\end{titlepage}
\setcounter{equation}{0}

\begin{centering}
\section{Introduction}
\end{centering}
\noindent The AdS/CFT correspondence \cite{malda} provides a
conceptual framework for consistently encoding the geometry of Anti-de
Sitter and special holonomy solutions of M-/string theory in a quantum
theory. Though the class of spacetimes to which it can be applied is
restricted, and unfortunately does not include FLRW cosmologies, it provides the only complete proposal extant
for the definition of a quantum theory of gravity. For the
prototypical example of $AdS_5\times S^5/\mathbb{R}^{10}$ and
$\mathcal{N}=4$ super Yang-Mills, the Maldacena conjecture is by now
approaching the status of proof \cite{beisert}, \cite{zarembo}. The literature
on the correspondence
is  enormous, from applications in pure mathematics to
phenomenological investigations. On the phenomenological front, much effort
has been devoted to extending 
the AdS/CFT correspondence from $\mathcal{N}=4$
super Yang-Mills to more realistic field theories
\cite{26} and even QCD itself \cite{kiritsis}, \cite{k1}. Also, recent
developments have raised the hope that we may soon be able to use
AdS/CFT to test M-/string theory in the lab \cite{sean}-\cite{rhic}. On the mathematical front, the motivation provided by the
AdS/CFT correspondence has stimulated spectacular progress in
differential geometry; early work on the correspondence showed
that there is a deep interplay
between Anti-de Sitter solutions of M-/string theory, singular special
holonomy manifolds and conformal field theories \cite{ks},
\cite{kleb}. This relationship has since been the topic of intense
investigation; a recent highlight has been the beautiful work on
Sasaki-Einstein geometry, toric Calabi-Yau three-folds and the
associated conformal field theories \cite{ypq}-\cite{6}. What has
become clear is that the geometry of a supersymmetric AdS/CFT dual
involves an Anti-de Sitter manifold, a singular special holonomy
manifold\footnote{With the obviously special non-singular exception of
  flat space.} and a supergravity solution which, in a sense that will
be made more precise, interpolates smoothly between them. This geometrical
relationship, between Anti-de Sitter manifolds and singular special
holonomy manifolds, in the context of the AdS/CFT correspondence in
M-theory, is the subject of this paper.

The canonical example of this relationship, from IIB, is that between conically
singular Calabi-Yau three-folds and Sasaki-Einstein $AdS_5$ solutions
of IIB supergravity. Each of these geometries, individually, is a supersymmetric
solution of IIB, preserving eight supercharges. Furthermore, the
manifolds may be superimposed\footnote{Because, with a suitable ansatz
  including both, the supergravity field equations linearise.} to
obtain another supersymmetric solution of IIB, admitting four
supersymmetries. This interpolating solution - the supergravity
description of D3 branes at a conical Calabi-Yau singularity - has
metric
\bea
\dd s^2=\left(\mbox{A}+\frac{\mbox{B}}{r^4}\right)^{-1/2}\dd
s^2(\mathbb{R}^{1,3})+\left(\mbox{A}+\frac{\mbox{B}}{r^4}\right)^{1/2}\Big(\dd
r^2+r^2\dd s^2(\mbox{SE}_5)\Big),
\eea
for constants A, B and a Sasaki-Einstein five-metric $\dd
s^2(\mbox{SE}_5)$. Setting B=0 gives the IIB solution
$\mathbb{R}^{1,3}\times \mbox{CY}_3$, while setting A=0 gives the solution
$\mbox{AdS}_5\times \mbox{SE}_5$. For positive A, B, the solution is globally
smooth, and contains two distinct asymptotic regions: a spacelike
infinity where the metric asymptotes to that of the Calabi-Yau, and
an internal spacelike infinity, where the metric asymptotes to that of
the Anti-de Sitter, on an event horizon at infinite proper
distance. The causal structure of these solutions is discussed in
detail in \cite{gibbons}. The Calabi-Yau singularity is excised in the
interpolating solution, and removed to infinity; an important feature
of the interpolating solution is that it admits a globally-defined
$SU(3)$ structure.

The AdS/CFT correspondence tells us how to perform this geometrical
interpolation in a quantum framework. Open string theory on the
singular Calabi-Yau reduces, at low energies, to a conformally
invariant quiver gauge theory, at weak 't Hooft coupling. This is the
low-energy effective field theory on the world-volume of a stack of probe D3
branes located at the singularity. The gauge theory encodes the toric
data of the Calabi-Yau. The same quiver gauge theory, at
strong 't Hooft coupling, is identical to IIB string theory on the
$\mbox{AdS}_5\times\mbox{SE}_5$; by the AdS/CFT dictionary, the CFT
also encodes the Sasaki-Einstein data of the $AdS$ solution. Clearly, it
can only do this for both the Calabi-Yau and the $AdS_5$ if their
geometry is intimately related. In the classical regime, this
relationship is provided by the interpolating solution. In the quantum
regime, the relationship is provided by the CFT itself; the
interpolation parameter is the 't Hooft coupling. In effect, the CFT
is telling us how to cut out the Calabi-Yau singularity quantum
gravitationally, and replace it with an event horizon with the
geometry of Anti-de Sitter. 

The correspondence is best understood for branes at conical
singularities of special holonomy manifolds. However, starting from
the work of Maldacena and Nu\~{n}ez \cite{mn}, many supersymmetric
$AdS$ solutions of M-/string theory have been
discovered,\cite{ach}-\cite{toni2}, \cite{ypq}, which cannot
be interpreted as coming from a stack of branes at a conical
singularity. Instead, they have been interpreted as the near-horizon
limits of the supergravity description of branes wrapped on calibrated
cyles of special holonomy manifolds. The CFT dual of the
$AdS$/special holonomy manifolds is the low-energy effective theory on
the unwrapped worldvolume directions of the branes. A
brane, heuristically envisioned as a hypersurface in spacetime, can
wrap a calibrated cycle in a special holonomy manifold, while
preserving supersymmetry. A heuristic physical argument as to why this is
possible is that a calibrated cycle is volume-minimising in its
homology class; as a probe brane has a tension, it will always try to
contract, and so a wrapped probe brane is only stable if it wraps a
minimal cycle. The supergravity description of a stack of wrapped branes, by
analogy with that of branes at conical singularities, should be a
supergravity solution which smoothly interpolates between a special
holonomy manifold with an appropriate calibrated cycle, and an event
horizon with Anti-de Sitter geometry. As the notion of an
interpolating solution is central to this paper, a more careful definition
of what is meant by these words will now be given.\\\\
\paragraph{Definition 1} Let $\mathcal{M}_{AdS}$ be a d-dimensional manifold
admiting a warped-product $AdS$ metric $\mbox{g}_{AdS}$, that, together
with a matter content $\mbox{F}_{AdS}$, gives a supersymmetric
solution of a supergravity theory in d dimensions. Let $\mathcal{M}_{SH}$ be a
d-dimensional manifold admitting a special holonomy metric
$\mbox{g}_{SH}$, which gives a supersymmetric vacuum solution of the
supergravity with holonomy $G\subset Spin(d-1)$. Let
$\mathcal{M}_{I}$ be a d-dimensional manifold admitting a globally-defined
G-structure, together with a metric $\mbox{g}_{I}$ and
a matter content $\mbox{F}_{I}$ that give a supersymmetric
solution of the supergravity. Then we say that
$(\mathcal{M}_{I},\mbox{g}_{I}, \mbox{F}_{I})$ is an {\it
  interpolating solution} if for all $\e,\zeta>0$, there exist open
sets $O_{AdS}\subset\mathcal{M}_{AdS},$
$O_{I},O'_{I}\subset\mathcal{M}_{I}$,
$O_{SH}\subset\mathcal{M}_{SH}$, such that for all points $p_{AdS}\in
O_{AdS}$, $p_{I}\in O_{I}$, $p'_{I}\in O'_{I}$,
$p_{SH}\in O_{SH}$,
\bea
|\mbox{g}_{AdS}(p_{AdS})-\mbox{g}_{I}(p_{I})|<\e,\;\;\;|\mbox{g}_{SH}(p_{SH})-\mbox{g}_{I}(p'_{I})|<\zeta.
\eea

\noindent We also define the following useful pieces of vocabulary:

\paragraph{Definition 2} If for a given pair
$(\mathcal{M}_{AdS},\mbox{g}_{AdS},\mbox{F}_{AdS})$,
$(\mathcal{M}_{SH},\mbox{g}_{SH},\mbox{F}_{SH})$, there exists an
interpolating solution, then we say that $\mathcal{M}_{SH}$ is a
{\it special holonomy interpolation} of $\mathcal{M}_{AdS}$ and that
$\mathcal{M}_{AdS}$ is an {\it Anti-de Sitter interpolation} of
$\mathcal{M}_{SH}$. Collectively, we refer to
$(\mathcal{M}_{AdS},\mbox{g}_{AdS},\mbox{F}_{AdS})$ and
$(\mathcal{M}_{SH},\mbox{g}_{SH},\mbox{F}_{SH})$  as an {\it
  interpolating pair}.\\

\noindent The objective of this paper is to derive candidate special
holonomy interpolations of some of the wrapped fivebrane near-horizon
limit $AdS$ solutions of
\cite{ach}-\cite{jk}. In \cite{me}, candidate special holonomy
interpolations of the $AdS_5$ M-theory solutions of \cite{mn} were
derived. These $AdS$ solutions describe the near-horizon limit of
fivebranes wrapped on K\"{a}hler two-cycles in Calabi-Yau two-folds
and three-folds. As these results fit nicely into the more extensive
picture presented here, they will be reviewed briefly below. The new
special holonomy metrics that will be derived here are candidate
interpolations of: the $AdS_3$ solution of \cite{gkw}, describing the
near-horizon limit of fivebranes wrapped on a K\"{a}hler four-cycle in
a four-fold; the $AdS_4$ solution of \cite{sezgin}, interpreted in
\cite{gkw} as the near-horizon limit of fivebranes on a special
lagrangian (SLAG)
three-cycle in a three-fold; the $AdS_3$ solution of \cite{gkw}, for
fivebranes on a SLAG four-cycle in a four-fold; the $AdS_4$ solution
of \cite{ach}, for fivebranes on an associative three-cycle in a $G_2$
manifold; the $AdS_3$ solution of \cite{gkw}, for fivebranes on a
co-associative four-cycle in a $G_2$ manifold; the $AdS_3$ solution of
\cite{jk}, for fivebranes on a complex lagrangian (CLAG) four-cycle in
an $Sp(2)$ manifold; and the $AdS_3$ solution of \cite{gkw}, for
fivebranes on a Cayley four-cycle in a $Spin(7)$ manifold. This paper
therefore provides one candidate interpolating pair for every type of cycle
on which M-theory fivebranes can wrap, in all manifolds of dimension
less than ten with irreducible holonomy,
with the exception of K\"{a}hler four-cycles in three-folds and
quaternionic K\"{a}hler four-cycles in 
$Sp(2)$ manifolds, for which no $AdS$ solutions are known to the author.

No interpolating solutions of eleven-dimensional supergravity which
describe wrapped branes are known. However, based on various symmetry
and supersymmetry arguments, the differential equations they satisfy
are known, for all types of calibrated cycles in all special holonomy manifolds
that play a r$\hat{\mbox{o}}$le in M-theory. These equations will be
called the wrapped brane equations; there is an extensive
literature on their derivation \cite{smith}-\cite{pau}; the most
general results are those of \cite{wrap}-\cite{pau}. The key point
that will be exploited here is that {\it both} members of an
interpolating pair should individually be a solution of the wrapped brane
equations, with a suitable ansatz for the interpolating solution. This
is just like what happens for an interpolating solution associated to
a conical special holonomy manifold.

One of the many important results of \cite{ypq} was to show how any
$AdS_5$ solution of M-theory, coming from fivebranes on a K\"{a}hler
two-cycle in a three-fold, satisfies the appropriate wrapped brane
equations. The canonical frame of the $AdS_5$ solutions, defined by
their eight Killing spinors, admits an $SU(2)$ structure. The $AdS_5$
solutions may also be re-written in such a way that the canonical
$AdS_5$ frame is obscured, but a canonical $\mathbb{R}^{1,3}$ frame is
made manifest. This frame admits an $SU(3)$ structure, and is defined
by {\it half} the Killing spinors of the $AdS_5$ solution. And it is
this Minkowski $SU(3)$ structure which satisfies the wrapped brane
equations. By definition, any interpolating solution describing
fivebranes on a K\"{a}hler two-cycle in a three-fold admits a
globally-defined $SU(3)$ structure; this structure smoothly matches on
to the $SU(3)$ structure of the Calabi-Yau and also to the canonical $SU(3)$
structure of the $AdS_5$ solution. This construction has since been
systematically extended to all calibrated cycles in manifolds with
irreducible holonomy of relevance to M-theory in \cite{wrap}, \cite{eoin}, \cite{pau}, and,
starting from the wrapped brane equations, has been used to classify
(ie, derive the differential equation satisfied by) all supersymmetric
$AdS$ solutions of M-theory which have a wrapped-brane origin.  

The strategy used here to construct candidate special holonomy
interpolations of the $AdS$ solutions is therefore the following. We
first construct the canonical Minkowski frames and structures of the
$AdS$ solutions, which satisfy the appropriate wrapped brane
equations. We then use these as a guide to formulating a suitable ansatz for an
interpolating solution. It is then a (reasonably) straightforward
matter to determine the most general special holonomy solution of the
$AdS$-inspired ansatz for the interpolating solution. In each case,
the special holonomy metric thus obtained is the proposed
interpolation of the $AdS$ solution. No attempt has been made to
determine the interpolating solutions themselves. It is therefore a
matter of conjecture whether the special holonomy metrics obtained are
indeed interpolations of the $AdS$ solutions. However the results
are sufficiently striking that it is reasonable to believe that for
the proposed interpolating pairs an interpolating solution does indeed
exist. 

As an illustration of this procedure, consider the results of
\cite{me} for the proposed interpolation of the $\mathcal{N}=2$
$AdS_5$ solution of \cite{mn}, describing the near-horizon limit of
fivebranes on a K\"{a}hler two-cycle in a two-fold. When re-written in
the canonical Minkowski frame, the $AdS$ solution is of the form
\bea
\dd s^2&=&L^{-1}\Big[\dd s^2(\mathbb{R}^{1,3})+\frac{F}{2}\dd
s^2(H^2)\Big]+L^2\Big[F^{-1}\Big(\dd u^2+u^2(\dd\psi-P)^2\Big)+\dd t^2+t^2\dd
s^2(S^2)\Big],\nn&&
\eea
where\footnote{Here, and throughout, $\dd s^2(AdS_n)$, $\dd s^2(H^n)$,
  $\dd s^2(S^n)$, denote the maximally symmetric Einstein metrics on
  n-dimensional $AdS$ manifolds, n-hyperboloids or n-spheres
  with unit radius of curvature, respectively. The cartesian metric on
  flat space will be denoted by $\dd s^2(\mathbb{R}^n)$. The volume
  form on a unit n-hyperboloid or n-sphere will be denoted by
  $\mbox{Vol}[H^n]$, $\mbox{Vol}[S^n]$, respectively.}
$\dd P=\mbox{Vol}[H^2]$, the period of $\psi$ is $2\pi$  and $F,L$ are known functions of the
coordinates $u$ and $t$. The ansatz for the interpolating solution is
then simply that $F,L$ are allowed to be arbitrary functions of
$u,t$. The most general special holonomy solution with this ansatz is
\bea
\dd s^2=\dd s^2(\mathbb{R}^{1,6})+\dd s^2(\mathcal{N}_{\tau}),
\eea
where, up to an overall scale,
\bea\label{kkj}
\dd s^2(\mathcal{N}_{\tau})=\frac{R^2}{4}\left[\dd
  s^2(H^2)+\left(\frac{1}{R^4}-1\right)(\dd \psi-P)^2\right]+
  \left(\frac{1}{R^4}-1\right)^{-1}\dd R^2.
\eea 
The range of $R$ is $R\in(0,1]$. At $R=1$, an $S^2$ degenerates
smoothly, and a $H^2$ bolt stabilises. At $R=0$, the metric is
singular, where the K\"{a}hler $H^2$ cycle degenerates. In the
probe-brane picture, the fivebranes should be thought of as wrapping
the $H^2$ at the singularity. Otherwise, they can always decrease
their worldvolume by moving to smaller $R$. This
incomplete special holonomy metric is to be compared with the
Eguchi-Hanson metric \cite{eguchi}, which is
\bea
\dd s^2(\mbox{EH})=\frac{R^2}{4}\left[\dd
  s^2(S^2)+\left(1-\frac{1}{R^4}\right)(\dd \psi-P)^2\right]+
  \left(1-\frac{1}{R^4}\right)^{-1}\dd R^2,
\eea  
where now $\dd P=\mbox{Vol}[S^2]$. As is well known, this metric is
complete in the range $R\in[1,\infty)$. At $R=1$, an
$S^2$ degenerates smoothly and a K\"{a}hler $S^2$ bolt stabilises.
 
In every case, the conjectured special holonomy interpolations of the
$AdS$ solutions derived in this paper are singular, and they have
exactly the same relationship with known complete special holonomy
metrics as that of \eqref{kkj} with Eguchi-Hanson. To make the pattern
clear, it worth quoting one more example now. The conjectured special
holonomy interpolation of the $AdS_3$ solution of \cite{gkw} for
fivebranes on a Cayley four-cycle in a $Spin(7)$ manifold is
\bea
\dd s^2=\dd s^2(\mathbb{R}^{1,2})+\dd s^2(\mathcal{N}_{\tau}),
\eea
where, up to an overall scale,
\bea
\dd s^2(\mathcal{N}_{\tau})=\frac{9}{20}R^2\dd
s^2(H^4)+\frac{36}{100}R^2\left(\frac{1}{R^{10/3}}-1\right)\mbox{D}Y^a\mbox{D}Y^a+\left(\frac{1}{R^{10/3}}-1\right)^{-1}\dd
R^2,
\eea
where the $Y^a$ are constrained coordinates on an $S^3$ and D will
be defined later. The range of $R$ is $R\in(0,1]$; at $R=1$ the $S^3$
degenerates smoothly and a $H^4$ bolt stabilises. At $R=0$ the metric
is singular where the $H^4$ Cayley four-cycle degenerates. This metric
is to be compared with the $Spin(7)$ metric on an $\mathbb{R}^4$
bundle over $S^4$, first found by Bryant and Salamon \cite{bryant} and
later independently by Gibbons, Page and Pope \cite{pope}:
\bea
\dd s^2(\mbox{BSGPP})=\frac{9}{20}R^2\dd
s^2(S^4)+\frac{36}{100}R^2\left(1-\frac{1}{R^{10/3}}\right)\mbox{D}Y^a\mbox{D}Y^a+\left(1-\frac{1}{R^{10/3}}\right)^{-1}\dd
R^2,
\eea
This metric is complete in the range $R\in[1,\infty)$; at $R=1$ an
$S^4$ degenerates smoothly and a Cayley $S^4$ bolt stabilises.

This relationship with known complete special holonomy metrics is a
universal feature of all the proposed special holonomy 
interpolations of this paper. As this series of incomplete special
holonomy metrics has so many
features in common, they will be given a collective name, the 
$\mathcal{N}_{\tau}$ {\it series}. Though they have been derived here
from the $AdS$ M-theory solutions {\it ab initio}, they may be
obtained in a much simpler way {\it a posteriori}, by analytic
continuation of known complete metrics\footnote{The
  $\mathcal{N}_{\tau}$ metrics have almost
  certainly been found before, though because they are incomplete,
  they have been presumably been rejected hitherto as pathological and
  uninteresting. What now makes them interesting is their
  interpretation as special holonomy interpolations of $AdS$
  solutions, for which their incompleteness is probably a
  pre-requisite: see conjecture 2 below.}.  In every case, they may be
obtained from a known complete metric with a radial coordinate of
semi-infinite range, at the endpoint of which an $S^m$ degenerates and
a calibrated $S^n$ (or, as appropriate, $\mathbb{CP}^2$) cycle
stabilises. The $\mathcal{N}_{\tau}$ series is obtained by changing
the sign of the scalar curvature of the bolt and analytically
continuing the dependence of the metric on the radial coordinate. This
generates a special holonomy metric with a ``radial'' coordinate of
finite range, with a smoothly degenerating $S^m$ and a stabilised
$H^n$ (or Bergman) bolt at one endpoint, and a singular degeneration at the
other. For the Calabi-Yau $\mathcal{N}_{\tau}$ with K\"{a}hler cycles in
three-folds and four-folds, the analogous known metrics are the
resolved conifold of \cite{cand}, \cite{pando}, and its
four-fold analogue (see \cite{tseytlin} for useful additional
background on the resolved conifold). For the Calabi-Yau $\mathcal{N}_{\tau}$ with SLAG
cycles, the analogous known metrics are the Stenzel metrics
\cite{stenzel} (see \cite{cvetic}, \cite{cvetic1} for useful
background on the Stenzel metrics). The Stenzel two-fold metric
coincides with Eguchi-Hanson, and the Stenzel three-fold metric
coincides with the deformed conifold metric of \cite{cand} (see
\cite{ohta}, \cite{tseytlin} for additional background on the deformed
conifold). For the $G_2$ $\mathcal{N}_{\tau}$ metrics with co-associative
cycles, the analogous known metrics are the BSGPP metrics
\cite{bryant}, \cite{pope} on $\mathbb{R}^3$ bundles over $S^4$ or
$\mathbb{CP}^2$. For the  $G_2$ $\mathcal{N}_{\tau}$ metric with an associative
cycle, the analogous known metric is the BSGPP metric
\cite{bryant}, \cite{pope} on an $\mathbb{R}^4$ bundle over
$S^3$. See \cite{cvetic2}, \cite{gukov}, \cite{cvetic1} for
more background on the complete $G_2$ metrics. For the $Sp(2)$ $\mathcal{N}_{\tau}$
metric with a CLAG cycle, the analogous known metric is the Calabi
metric on $T^*\mathbb{CP}^2$ \cite{calabi}; the Calabi metric is the
unique complete regular hyper-K\"{a}hler eight-manifold of co-homogeneity one
\cite{swann}; for further background on the
Calabi metric, see
\cite{cvetic3}. Finally, for the $Spin(7)$ $\mathcal{N}_{\tau}$ metric
with a Cayley four-cycle, we have seen that the analogous known metric
is the BSGPP metric on an $\mathbb{R}^4$ bundle over $S^4$;
see \cite{cvetic2}, \cite{gukov}, \cite{cvetic1} for more details.   

What is most striking about the conjectured special holonomy
interpolations obtained here is that they are all singular. As occurs
in the conical context, the expectation is that the singularity of the
special holonomy manifold is excised in the interpolating solution,
and that the conformal dual of the geometry gives a quantum
gravitational definition of this process. If this is correct, then a
singularity of the special holonomy manifold is an essential
ingredient of the geometry of AdS/CFT. It would also explain a
hitherto rather puzzling feature of the $AdS$ solutions studied
here, all of which were originally constructed in gauged
supergravity. While for the $\mathcal{N}_{\tau}$ series it is possible to
obtain the known special holonomy manifolds by replacing the $H^n$ factors
with $S^n$ factors, for their $AdS$ interpolations this
does not seem to be possible; the $AdS$ solutions exist only for
hyperbolic cycles. This makes sense if an AdS/CFT dual can exist only
for a singular special holonomy manifold; otherwise, if $AdS$
solutions like those studied here, but with $S^n$ cycles, existed,
their special holonomy interpolations would be non-singular. Another
way of saying this is that it seems that a conformal field theory can
be associated to the singular $\mathcal{N}_{\tau}$ series of special
holonomy metrics, but {\it not} to their non-singular known
analogues. If this idea is correct, it means that what the AdS/CFT
correspondence is ultimately describing is the quantum gravity of singularity
resolution for special holonomy manifolds. We formalise the geometry
of this idea in the following two conjectures.     

\paragraph{Conjecture 1} Every supersymmetric Anti-de Sitter solution
of M-/string theory admits a special holonomy interpolation.

\paragraph{Conjecture 2} With the exception of flat space, the metric
on every special holonomy manifold admitting an Anti-de Sitter
interpolation is incomplete. \\\\   

The organisation of the remainder of this paper is as follows. In
section two, as useful introductory material, we will review the
relationship between the canonical $AdS$ and Minkowski frames for
$AdS$ solutions, how to
pass from one to the other by means of a frame rotation, and the
relationship between the $AdS$ and wrapped brane structures. In
section three, we will derive the conjectured special holonomy interpolations
of $AdS$ solutions for fivebranes wrapped on cycles in Calabi-Yau
manifolds. Section four is devoted to the proposed $Sp(2)$ interpolating pair,
section five to the $G_2$ interpolating pairs and section six to the $Spin(7)$
interpolating pair. In section seven we conclude and discuss interesting future
directions.  

\begin{centering}
\section{Canonical Minkowski frames for AdS manifolds}
\end{centering}

\noindent In this section we will review how the canonical $AdS$ frame defined
by all the Killing spinors of a supersymmetric $AdS$ solution is
related to its canonical Minkowski frame defined by half its Killing
spinors; for more details, the reader is referred to \cite{ypq},
\cite{wrap}-\cite{pau}.  The canonical Minkowski structure of an $AdS$ solution
is the one which can match on to the 
G-structure of an interpolating solution. This phenomenon - the
matching of the structure defined by half the supersymmetries of the
$AdS$ manifold to that of an interpolating solution - is another, more
precise way of stating the familiar feature of supersymmetry doubling
in the near-horizon limit of a supergravity brane solution. 

We will in fact distinguish two cases, which will be discussed
seperately. The $AdS$ solutions we study for fivebranes on cycles in
manifolds of $SU(2)$, $SU(3)$ or $G_2$ holonomy have purely magnetic
fluxes. This means that no membranes are present in the
geometry. However, the $AdS$ solutions for fivebranes on four-cycles
in eight-manifolds ($Spin(7)$, $SU(4)$ or $Sp(2)$ holonomies) have both
electric and magnetic fluxes. In probe-brane language, we can think of
a stack of 
fivebranes wrapped a four-cycle in the eight-manifold. We also have a
stack of membranes extended in the three overall transverse directions
to the eight-manifold. The membrane stack intersects the fivebrane
stack in a string; the low-energy effective field theory on the string
worldvolume is then the two-dimensional dual of the $AdS_3$ solutions
that come from these geometries. The presence of the membranes
complicates the relationship of the $AdS$ and Minkowski frames a
little, so first we will discuss the case of fivebranes alone, and
purely magnetic fluxes.\\\\

\begin{centering}
\subsection{AdS spacetimes from fivebranes on cycles in $SU(2)$,
  $SU(3)$ and $G_2$ manifolds}
\end{centering}

\noindent The metric of an interpolating solution describing a stack of
fivebranes wrapped on a calibrated cycle in a Calabi-Yau two- or
three-fold, or a $G_2$ manifold, takes the form
\bea\label{kop}
\dd s^2=L^{-1}\dd s^2(\mathbb{R}^{1,p})+\dd s^2(\mathcal{M}_q)+L^2\Big(\dd
t^2+t^2\dd s^2\left(S^{10-p-q}\right)\Big),
\eea
where $\mathcal{M}_q$ admits a globally-defined $SU(2)$, $SU(3)$ or
$G_2$ structure respectively. The Minkowski isometries are isometries
of the full solution, and the flux has no components
along the Minkowski directions. The dimensionality of $\mathcal{M}_q$ is
$q=4,6,7$, respectively. The dimensionality of the unwrapped fivebrane
worldvolume is $p+1$, so $p=3$ for a K\"{a}hler two-cycle, $p=2$ for a
SLAG or associative three-cycle, and $p=1$ for a co-associative
four-cyle. The intrinsic torsion of the G-structure on $\mathcal{M}_q$
must satisfy certain conditions, implied by supersymmetry and the
four-form Bianchi identity. These conditions are what are called the
wrapped brane equations; they will be given for each case below, and
need not concern us now. For more details, the reader is referred to
\cite{wrap}. 

Our interest here is how to obtain a warped product $AdS$ metric from
the wrapped-brane metric \eqref{kop}, and vice versa. The first step is to recognise
that every warped-product $AdS_{p+2}$ metric, written in Poincar\'{e}
coordinates, may be thought of as a special case of a warped
$\mathbb{R}^{1,p}$ metric. If the $AdS$ warp factor is denoted by
$\lambda$, and is independent of the $AdS$ coordinates, then
\bea
\l^{-1}\dd s^2(AdS_{p+2})=\l^{-1}[e^{-2r}\dd s^2(\mathbb{R}^{1,p})+\dd
r^2].
\eea
Therefore our first step is to identify $L=\l e^{2r}$ in \eqref{kop},
with $r$ the $AdS$ radial coordinate. The next step is to pick out the
$AdS$ radial direction $\hat{r}=\l^{-1/2}\dd r$ from the space
transverse to the $\mathbb{R}^{1,p}$ factor in \eqref{kop}. In the
cases of interest to us, the $AdS$ radial direction is a linear
combination of the radial direction $\hat{v}=L\dd t$ on the overall transverse
space, and a radial direction in $\mathcal{M}_q$, transverse to the
wrapped cycle. We denote this radial basis one-form on $\mathcal{M}_q$
by $\hat{u}$. Thus we can obtain the $AdS$ radial basis one-form by
a local rotation of the frame of \eqref{kop}:
\bea
\hat{r}=\sin\theta\hat{u}+\cos\theta\hat{v},
\eea
for some local angle $\theta$ which we take to be independent of
$r$. Denoting the orthogonal linear combination in the $AdS$ frame by
$\hat{\rho}$, we have
\bea
\hat{\rho}=\cos\theta\hat{u}-\sin\theta\hat{v}.
\eea
Now, imposing closure of $\dd t$ and $r$-independence of $\theta$, we
get
\bea
\hat{\rho}=\frac{\l}{2\sin\theta}\dd(\l^{-3/2}\cos\theta).
\eea
Defining a coordinate $\rho$ for the $AdS$ frame according to
$\rho=\l^{-3/2}\cos\theta$, we get   
\bea
t&=&-\frac{\r}{2}e^{-2r},\nn
\hat{\rho}&=&\frac{\l}{2\sqrt{1-\l^3\r^2}}\dd\rho.
\eea
Finally, we impose that the metric on the space tranverse to the $AdS$
factor is independent of the $AdS$ radial coordinate, and (in deriving
the $AdS$ supersymmetry conditions from the wrapped brane equations)
that the flux has no components along the $AdS$ radial direction. Thus
we obtain the (for our purposes) general $AdS_{p+2}$ metric contained
in \eqref{kop}:
\bea\label{kopp}
\dd s^2=\l^{-1}\left[\dd
  s^2(AdS_{p+2})+\frac{\l^3}{4}\left(\frac{\dd\rho^2}{1-\l^3\r^2}+\r^2\dd
    s^2\left(S^{10-p-q}\right)\right)\right]+\dd s^2(\mathcal{M}_{q-1}),
\eea
where $\dd s^2(\mathcal{M}_{q-1})$ is defined by
\bea
\dd s^2(\mathcal{M}_q)=\dd
s^2(\mathcal{M}_{q-1})+\hat{u}\otimes\hat{u}.
\eea
In addition, we have
\bea\label{uu}
\hat{u}=\l\left(\sqrt{\frac{1-\l^3\r^2}{\l^3}}\dd
  r+\sqrt{\frac{\l^3}{1-\l^3\r^2}}\frac{\r}{2}\dd\r\right).
\eea
Since in general we know the relationship between the Minkowski-frame
coordinate $t$ and the $AdS$ frame coordinates $r,\r$, when we know
$\l$ explicitly for a particular solution, we can integrate \eqref{uu}
to find an explicit coordinatisation of the $AdS$ solution in the
Minkowski frame. Thus we can pass freely from one frame to the other,
for any explicit solution.

Having discussed the relationship of the frames, let us now discuss
the relationship between the structures. Since, in passing from
\eqref{kop} to \eqref{kopp} we pick out a preferred direction on
$\mathcal{M}_{q}$, the G-structure of \eqref{kop} on $\mathcal{M}_q$
is reduced to a $G'$ structure on $\mathcal{M}_{q-1}$ in
\eqref{kopp}. For $q=4$, the $SU(2)$ structure on $\mathcal{M}_4$ is
reduced to an identity structure on $\mathcal{M}_3$; the $SU(2)$ forms
on $\mathcal{M}_4$ decompose according to
\bea
J_4&=&e^1\w e^2+e^3\w\hat{u},\\
\Omega_4&=&(e^1+ie^2)\w(e^3+i\hat{u}),
\eea
with 
\bea
\dd s^2(\mathcal{M}_4)=e^1\otimes e^1+e^2\otimes e^2+e^3\otimes
e^3+\hat{u}\otimes \hat{u}.
\eea
For $q=6$, the $SU(3)$ structure on $\mathcal{M}_6$ reduces to an
$SU(2)$ structure on $\mathcal{M}_5$; the $SU(3)$ structure forms
decompose according to
\bea
J_6&=&J_4+e^5\w \hat{u},\nn
\Omega_6&=&\Omega_4\w(e^5+i\hat{u}),
\eea
with
\bea
\dd s^2(\mathcal{M}_6)=\dd
s^2(\mathcal{M}_5)+\hat{u}\otimes\hat{u}=\dd
s^2(\mathcal{M}_4)+e^5\otimes e^5+\hat{u}\otimes\hat{u},
\eea
and the $SU(2)$ structure of the $AdS$ frame is defined on
$\mathcal{M}_4$. For $q=7$, the $G_2$ structure on $\mathcal{M}_7$ reduces to an
$SU(3)$ structure on $\mathcal{M}_6$; the $G_2$ structure forms decompose
according to
\bea
\Phi&=&J_6\w\hat{u}-\mbox{Im}\Omega_6,\nn
\Upsilon&=&\frac{1}{2}J_6\w J_6+\mbox{Re}\Omega_6\w\hat{u},
\eea
with
\bea
\dd s^2(\mathcal{M}_7)=\dd s^2(\mathcal{M}_6)+\hat{u}\otimes\hat{u},
\eea
and the $SU(3)$ structure of the $AdS$ frame is defined on
$\mathcal{M}_6$.\\\\

\begin{centering}
\subsection{AdS spacetimes from fivebranes on four-cycles in eight-manifolds of Spin(7),
  SU(4) or Sp(2) holonomy}
\end{centering}

\noindent As discussed above, because of the presence of non-zero electric flux
for $AdS_3$ solutions from fivebranes on four-cycles in
eight-manifolds, the relationship between the canonical $AdS$ and
Minkowski frames of the $AdS$ solutions is a little more
complicated. These systems are the subject of \cite{pau}, to which
the reader is referred for more details\footnote{In \cite{pau}, 
  somewhat more general wrapped brane metrics were considered than
  those of this discussion. However the discussion of this section is sufficiently
  general for the applications of interest in this paper.}. The metric of an interpolating solution describing a
stack of fivebranes wrapped on a four-cycle in an eight-manifold, with
a stack of membranes extended in the transverse directions, takes the
form
\bea\label{oi}
\dd s^2=L^{-1}\dd s^2(\mathbb{R}^{1,1})+\dd s^2(\mathcal{M}_8)+C^2\dd
t^2.
\eea
Again, the Minkowski isometries are isometries of the full solution,
the electric flux contains a factor proportional to the Minkowski
volume form, and the magnetic flux has no components along the
Minkowski directions. The Minkowski directions represent the unwrapped
fivebrane worldvolume directions; the membranes extend in these
directions and also along $\dd t$. Note that in this case the warp
factor of the overall transverse space (the $\mathbb{R}$ coordinatised
by $t$) is independent of the Minkowski warp factor. The global
G-structure is defined on $\mathcal{M}_8$; the structure group is
$Spin(7)$, $SU(4)$ or $Sp(2)$, as appropriate. Again, supersymmetry,
the four-form Bianchi identity, and now, the four-form field equation
imply restrictions on the intrinsic torsion of the global
G-structure. These equations, the wrapped brane equations for these
systems, are given in \cite{pau}.

To obtain an $AdS_3$ metric from \eqref{oi}, we again require that
that $L=\l e^{2r}$, with $r$ the $AdS$ radial coordinate and $\l$ the
$AdS$ warp factor, which we require to be independent of the $AdS$
coordinates. As before, we must now pick out the $AdS$ radial
direction $\hat{r}=\l^{-1/2}\dd r$ from the space transverse to the
Minkowski factor. In the generic case of interest to us, the $AdS$
radial direction is a linear combination of the overall
transverse direction $e^9=C\dd t$ and a radial direction in
$\mathcal{M}_8$ transverse to the cycle that we denote by $e^8$. Thus,
as before, we write the frame rotation relating the Minkowski and
$AdS$ frames as
\bea
\hat{r}&=&\sin\theta e^8+\cos\theta e^9,\nn
\hat{\rho}&=&\cos\theta e^8-\sin\theta e^9,
\eea
for a local rotation angle $\theta$ which we take to be independent of
the $AdS$ radial coordinate. Imposing $AdS$ isometries on the electric
and magnetic flux,
and requiring that the metric on the space transverse to the $AdS$
factor is independent of the $AdS$ coordinates, we find that we may
introduce an $AdS$ frame coordinate $\rho$ such that
\bea
\l^{-3/2}\cos\theta&=&f(\rho),\nn
\hat{\rho}&=&\frac{\l}{2\sqrt{1-\l^3f^2}}\dd \r,
\eea
for some arbitrary function $f(\rho)$. See \cite{pau} for a fuller
discussion of this point. Then the general $AdS$
metric contained in \eqref{oi} is
\bea
\dd s^2=\frac{1}{\l}\left[\dd s^2(AdS_3)+\frac{\l^3}{4(1-\l^3 f^2)}\dd
  \r^2\right]+\dd s^2(\mathcal{M}_7),
\eea
where $\dd s^2(\mathcal{M}_7)$ is defined by
\bea
\dd s^2(\mathcal{M}_8)=\dd s^2(\mathcal{M}_7)+ e^8\otimes e^8.
\eea
The basis one-forms of the Minkowski frame are given in terms of the
basis one-forms of the $AdS$ frame by
\bea
e^8&=&\l\left(\sqrt{\frac{1-\l^3 f^2}{\l^3}}\dd
  r+\sqrt{\frac{\l^3}{1-\l^3 f^2}}\frac{f}{2}\dd\r\right),\nn
C\dd t&=&\l f\dd r-\frac{1}{2}\l\dd\r.
\eea
For an explicit $AdS_3$ solution we know $\l$ and $f$ explicitly, and
so we can integrate these expressions to get an explicit coordinatisation
of the $AdS$ solution in the Minkowski frame. Thus we can freely pass
between the canonical $AdS$ and Minkowski frames for known $AdS$
solutions. 

As in the previous subsection, because we are picking out a preferred
direction on $\mathcal{M}_8$, the Minkowski-frame structure on
$\mathcal{M}_8$ is reduced, in the $AdS$ frame, to a structure on
$\mathcal{M}_7$. A $Spin(7)$ structure on $\mathcal{M}_8$ is reduced
to a $G_2$ structure on $\mathcal{M}_7$; the decomposition of the
Cayley four-form is
\bea
-\phi=\Upsilon+\Phi\w e^8.
\eea
An $SU(4)$ structure on $\mathcal{M}_8$ is reduced to an $SU(3)$
structure on $\mathcal{M}_7$. The decomposition of the $SU(4)$
structure forms is
\bea
J_8&=&J_6+e^7\w e^8,\nn
\Omega_8&=&\Omega_6\w(e^7+ie^8),
\eea
with
\bea
\dd s^2(\mathcal{M}_8)=\dd s^2(\mathcal{M}_7)+e^8\otimes e^8=\dd
s^2(\mathcal{M}_6)+e^7\otimes e^7+e^8\otimes e^8,
\eea
with the $SU(3)$ structure forms defined on $\dd
s^2(\mathcal{M}_6)$. Finally, an $Sp(2)$ structure on $\mathcal{M}_8$
reduces to an $SU(2)$ structure on $\dd s^2(\mathcal{M}_7)$. The
decomposition of the triplet of $Sp(2)$ almost complex structures
(which obey the algebra $J^AJ^B=-\delta^{AB}+\e^{ABC}J^C$, $A=1,2,3$)
under $SU(2)$ is
\bea
J^1&=&K^3+e^5\w e^6+e^7\w e^8,\nn
J^2&=&K^2-e^5\w e^7+e^6\w e^8,\nn
J^3&=&K^1+e^6\w e^7+e^5\w e^8,
\eea
with
\bea
\dd s^2(\mathcal{M}_8)=\dd s^2(\mathcal{M}_4)+e^5\otimes
e^5+e^6\otimes e^6+e^7\otimes e^7+e^8\otimes e^8,
\eea
and the $K^A$ are a triplet of self-dual $SU(2)$-invariant two-forms on $\mathcal{M}_4$,
which satisfy the algebra\footnote{The slightly eccentric labelling of
  the $SU(2)$ structure forms is chosen to coincide with an
  unfortunate conventional quirk of \cite{pau}.}
$K^AK^B=-\delta^{AB}-\e^{ABC}K^C.$ Having concluded the introductory
review, we now move on to the main results of the paper.

\begin{centering}
\section{Calabi-Yau interpolating pairs}
\end{centering}

\noindent In this section, we will give conjectured interpolating
pairs for fivebranes wrapped on calibrated cycles in Calabi-Yau
manifolds. First we will discuss K\"{a}hler cycles, then SLAG
cycles. In order to present a complete picture, we will summarise the
results of \cite{me} for K\"{a}hler two-cycles in two-folds and
three-folds. In the new cases, we will first present the pair, and
then give the derivation of the special holonomy interpolation from
the $AdS$ solution.\\\\

\begin{centering}
\subsection{K\"{a}hler cycles}
\end{centering}

\noindent In this subsection, the $AdS$ solutions for which we give a
conjectured special holonomy interpolation are: the half-BPS $AdS_5$
solution of \cite{mn}, describing the near-horizon limit of fivebranes
on a two-cycle in a two-fold; the quarter-BPS $AdS_5$ solution of
\cite{mn}, for a two-cycle in a three-fold; and the $AdS_3$ solution
of \cite{gkw}, admitting four Killing spinors, for a four-cycle in a
four-fold. The special holonomy interpolations of the first two cases
are derived in \cite{me}; here we will just describe the conjectured
pair. All the other pairs given in this paper are new, and their
derivation will be given.\\\\

\begin{centering}
\subsubsection{Two-fold}
\end{centering}

\paragraph{The conjectured interpolating pair} The metric of the half-BPS $AdS_5$
solution of \cite{mn} is given by
\bea\
\dd s^2&=&\frac{1}{\l}\left[\dd s^2(AdS_5)+\frac{1}{2}\dd
  s^2(H^2)+(1-\l^3\r^2)(\dd
  \psi-P)^2+\frac{\l^3}{4}\left(\frac{\dd\r^2}{1-\l^3\r^2}+\r^2\dd
    s^2(S^2)\right)\right],\nonumber
\eea
\bea
\l^3&=&\frac{8}{1+4\r^2},\nn
\eea
where $\dd P=\mbox{Vol}[H^2]$. The range of the coordinate
$\rho$, which without loss of generality we take to be non-negative,
is $\rho\in[0,1/2]$. At $\rho=0$, the R-symmetry $S^2$
degenerates smoothly\footnote{The R-symmetry of the dual theory is $SU(2)\times
  U(1)$.}. At
$\rho=1/2$, the R-symmetry $U(1)$, with coordinate
$\psi$, degenerates smoothly, provided that $\psi$ is identified with
period $2\pi$.\\

\noindent As discussed in the introduction, the conjectured special holonomy
interpolation of this manifold is 
\bea
\dd s^2(\mathcal{N}_{\tau})=\dd s^2(\mathbb{R}^{1,6})+\dd s^2(\mathcal{N}_{\tau}),
\eea
where, up to an overall scale,
\bea
\dd s^2(\mathcal{N}_{\tau})=\frac{R^2}{4}\left[\dd
  s^2(H^2)+\left(\frac{1}{R^4}-1\right)(\dd \psi-P)^2\right]+
  \left(\frac{1}{R^4}-1\right)^{-1}\dd R^2.
\eea 
The range of $R$ is $R\in(0,1]$. At $R=1$, an $S^2$ degenerates
smoothly, provided that $\psi$ has the same period as in the $AdS$
solution. At $R=0$, the metric is
singular, where the K\"{a}hler $H^2$ cycle degenerates.

\begin{centering}
\subsubsection{Three-fold}
\end{centering}

\paragraph{The conjectured interpolating pair} The metric of the
quarter-BPS $AdS_5$ solution of \cite{mn} is
\bea
\dd s^2&=&\frac{1}{\l}\left[\dd s^2(AdS_5)+\frac{1}{3}\dd
  s^2(H^2)+\frac{1}{9}(1-\l^3\r^2)\Big(\dd
  s^2(S^2)+(\dd\psi+P-P')^2\Big)+\frac{\l^3}{4(1-\l^3\r^2)}\dd\rho^2\right],\nonumber
\eea
\bea
\l&=&\frac{4}{4+\rho^2},
\eea
where now $\dd P=\mbox{Vol}[S^2],$ $\dd P'=\mbox{Vol}[H^2].$ This
time, the range of $\rho$ is $[-2/\sqrt{3},2/\sqrt{3}]$; at
$\rho=\pm2/\sqrt{3}$, an $S^3$ degenerates smoothly, provided that
$\psi$ is periodically identified with period $4\pi$.\\

\noindent The conjectured special holonomy interpolation of this manifold is
\bea
\dd s^2=\dd s^2(\mathbb{R}^{1,4})+\dd s^2(\mathcal{N}_{\tau}),
\eea
where, up to an overall scale,
\bea
\dd s^2(\mathcal{N}_{\tau})&=&\frac{1}{2}(1+\sin\xi)\dd
s^2(H^2)+\frac{\cos^2\xi}{2(1+\sin\xi)}\dd
s^2(S^2)+\frac{1}{\cos^2\xi}\Big(\dd
R^2+R^2(\dd\psi+P-P')^2\Big), \nonumber
\eea
\bea\label{218}
-\frac{1}{3}\sin^3\xi+\sin\xi=\frac{2}{3}-R^2.
\eea 
The range of $R$ is $R\in[0,2/\sqrt{3})$. At $R=0$ (corresponding to
$\xi=\pi/2$) an $S^3$ degenerates smoothly, provided that $\psi$ has
the same periodicity as for the $AdS$ coordinate. The metric is
singular at $R=2/\sqrt{3}$ (corresponding to $\xi=-\pi/2$) where the
K\"{a}hler $H^2$ cycle degenerates. This metric is the hyperbolic
analogue of the resolved conifold metric of \cite{cand}, \cite{pando}.  

\begin{centering}
\subsubsection{Four-folds}
\end{centering}

\paragraph{The interpolating pairs} This is the first new case we
encounter. A set of $AdS_3$ solutions was constructed by Gauntlett,
Kim and Waldram (GKW) in \cite{gkw}, that describe the near-horizon
limit of M5 branes on a K\"{a}hler
four-cycle in a Calabi-Yau four-fold, intersecting membranes extended
in the directions transverse to the four-fold. The $AdS$ solutions
admit four Killing spinors, and are as follows. The metrics are 
\bea
\dd s^2=\frac{1}{\l}\left[\dd s^2(AdS_3)+\frac{3}{4}\dd
  s^2(\mbox{KE}_4^-)+\frac{1}{4}(1-\l^3f^2)\Big(\dd s^2(S^2)+(\dd
    \psi+P+P')^2\Big)+\frac{\l^3}{4(1-\l^3f^2)}\dd\r^2\right],\nonumber
\eea
\bea
\l^3=\frac{9}{12+\r^2},\;\;f=\frac{2\r}{3}.
\eea
Here $\mbox{KE}_4^-$ is an arbitrary negative scalar curvature
K\"{a}hler-Einstein manifold, normalised such that the Ricci form
$\mathcal{R}_4$ is given by $\mathcal{R}_4=-\hat{J}_4$, with $\hat{J}_4$
the K\"{a}hler form of $\mbox{KE}_4^-$. In addition,
\bea
\dd P&=&\mbox{Vol}[S^2],\nn
\dd P'&=&\mathcal{R}_4.
\eea
The range of $\r$ is $\r\in[-2,2]$; at the end-points, an $S^3$
smoothly degenerates, provided that $\psi$ is periodically identified
with period $4\pi$. These manifolds admit an $SU(3)$ structure, which
was obtained in 
\cite{pau}, and will be given below (in somewhat more transparent coordinates), together with the magnetic flux
(the electric flux, which is irrelevant to the discussion, can be
obtained from \cite{gkw} or \cite{pau}).\\  

\noindent The conjectured special holonomy interpolation of these manifolds is
\bea
\dd s^2=\dd s^2(\mathbb{R}^{1,2})+\dd s^2(\mathcal{N}_{\tau}),
\eea
where, up to an overall scale,
\bea
\dd s^2(\mathcal{N}_{\tau})=\frac{1}{2}(1+\sin\xi)\dd
s^2(\mbox{KE}_4^-)+\frac{\cos^2\xi}{2(1+\sin\xi)}\dd
s^2(S^2)+\frac{1}{\cos^2\xi}\Big(\dd
R^2+R^2(\dd\psi+P-P')^2\Big), \nonumber
\eea
\bea
-\frac{1}{3}\sin^3\xi+\sin\xi=\frac{2}{3}-R^2.
\eea 
This is identical to the three-fold metric  of the previous
subsection, but with the $H^2$ replaced by a $\mbox{KE}_4^-$. It has
the same regularity properties, and is the hyperbolic analogue of the
four-fold resolved conifold. Now we will discuss its derivation.

\paragraph{The G-structure of the AdS solutions} First we will give the
$SU(3)$ structure of the $AdS$ solutions, defined by all four Killing spinors. Defining the frame
\bea
e^a&=&\sqrt{\frac{3}{4\l}}\hat{e}^a,\nn
e^5+ie^6&=&\frac{1}{2}\sqrt{1-\l^3f^2}e^{i\psi}(\dd\theta+i\sin\theta\dd\phi),\nn
e^7&=&\frac{1}{2}\sqrt{1-\l^3f^2}(\dd\psi+P+P'),
\eea
where $a=1,...,4$, the $\hat{e}^a$ furnish a basis for
$\mbox{KE}_4^-$, $\hat{J}_4=\hat{e}^{12}+\hat{e}^{34}$ and
$\hat{\Omega}_4=(\hat{e}^1+i\hat{e^2})(\hat{e}^3+i\hat{e}^4)$, the
$SU(3)$ structure is given by 
\bea
J_6&=&e^{12}+e^{34}+e^{56},\nn
\Omega_6&=&(e^1+ie^2)(e^3+ie^4)(e^5+ie^6).\nn
\eea
This structure is a solution of the torsion conditions of \cite{pau}
for the near-horizon limit of fivebranes on a K\"{a}hler four-cycle in
a four-fold, which are
\bea\label{eqn:k4jj}
\hat{\rho}\w\dd(\l^{-1}J_6\w J_6)&=&0,\\
\dd(\l^{-3/2}\sqrt{1-\l^3f^2}\mbox{Im}\Omega_6)&=&2\l^{-1}(e^7\w\mbox{Re}\Omega_6-\l^{3/2}f\hat{\r}\w\mbox{Im}\Omega_6),\\\label{eqn:k4de7}
J_6\lrcorner\;\dd
e^7&=&\frac{2\l^{1/2}}{\sqrt{1-\l^3f^2}}(1-\l^3f^2)-\l^{3/2}f\hat{\r}\lrcorner\;\dd\log\left(\frac{\l^3f}{1-\l^3f^2}\right).
\eea
In addition it is a solution of the Bianchi identity for the magnetic
flux, $\dd F_{\mbox{\small{mag}}}=0$, which in this case is not implied by the
torsion conditions. The magnetic flux is given by
\bea
F_{\mbox{\small{mag}}} &=&  \frac{\l^{3/2}}{\sqrt{1-\l^3f^2}}(\l^{3/2}f+\star_8)(\dd[\l^{-3/2}\sqrt{1-\l^3f^2}
J_6\w e^{7}
]-2\l^{-1} J_6\w J_6  ) + 2\l^{1/2} J_6\w  e^{7}\w \hat{\r},\nn
\eea
where $\star_8$ is the Hodge dual on the space transverse to the $AdS$
factor, with positive orientation defined with repect to
\bea
\mbox{Vol}=\frac{1}{3!}J_6\w J_6\w J_6\w e^7\w \hat{\rho}.
\eea

\paragraph{The $AdS$ solutions in the Minkowski frame} Now we use the
discussion of section 2 to frame-rotate the $AdS$ solutions to the
canonical Minkowski frame. Defining the coordinates
\bea
t&=&-\frac{1}{2}e^{-4r/3}\r,\nn
u&=&-\frac{1}{3}\sqrt{12-3\r^2}e^{-r},
\eea
the one-forms $e^8$, $e^9$ in the Minkowski frame are given by
\bea
e^8&=&\l e^r\dd u,\nn
e^9&=&\l e^{4r/3}\dd t,
\eea
and the metric in the Minkowski frame takes the form
\bea
\dd s^2&=&\frac{1}{H^{1/3}_{M5}H_{M2}^{2/3}}\dd
s^2(\mathbb{R}^{1,1})+\frac{H^{2/3}_{M5}}{H^{2/3}_{M2}}\dd
t^2+\frac{H^{1/3}_{M2}}{H^{1/3}_{M5}}\left[\frac{3}{4}F\dd
  s^2(\mbox{KE}_4^-)\right]\nn&&+H^{1/3}_{M2}H_{M5}^{2/3}\left[\frac{1}{F}\left(\dd
    u^2+\frac{u^2}{4}[\dd s^2(S^2)+(\dd \psi+P+P')^2]\right)\right],
\eea
where
\bea
H_{M5}&=&\l^3 e^{14r/3},\nn
H_{M2}&=&e^{2r/3},\nn
F&=&e^{4r/3}.
\eea
These three functions have been chosen so that the metric takes a form
reminiscent of the harmonic function superposition rule for
intersecting branes, in line with the probe brane picture. The
fivebrane worldvolume directions are the Minkowski and $\mbox{KE}_4^-$
directions; the membranes extend along the Minkowski and $t$
directions. Also $e^{2r}$ is given in terms of $t$ and $u$ by a
positive signature metric inducing root of the quartic
\bea
t^6e^{8r}  -\left(1-\frac{3}{4}u^2e^{2r}\right)^3=0.
\eea
The wrapped-brane $SU(4)$ structure of the $AdS_3$ solutions, defined
by two of their Killing spinors, is given
by
\bea
J_8&=&J_6+e^{7}\w e^8,\nn
\Omega_8&=&\Omega_6\w(e^7+ie^8).
\eea
By construction, this structure is a solution of the wrapped brane
equations for a K\"{a}hler four-cycle in a four-fold. These comprise
the torsion conditions \cite{oap2}, \cite{pau}
\bea
J_8\lrcorner\;\dd e^9&=&0,\nn
\dd(L^{-1}\mbox{Re}\Omega_8)&=&0,\nn
e^9\w[J_8\lrcorner \;\dd J_8-Le^9\lrcorner\;\dd(L^{-1}e^9)]&=&0,
\eea
and the Bianchi identity and field equation for the four-form, which
is given in the Minkowski frame in \cite{oap2}, \cite{pau}.

\paragraph{The conjectured Calabi-Yau interpolation} We now make the
following ansatz for an interpolating solution:
\bea
\dd s^2&=&\frac{1}{H^{1/3}_{M5}H_{M2}^{2/3}}\dd
s^2(\mathbb{R}^{1,1})+\frac{H^{2/3}_{M5}}{H^{2/3}_{M2}}\dd
t^2+\frac{H^{1/3}_{M2}}{H^{1/3}_{M5}}\left[\a^2F_1^2F_2^2\dd
  s^2(\mbox{KE}_4^-)\right]\nn&&+H^{1/3}_{M2}H_{M5}^{2/3}\left[\frac{1}{F_1^2}\left(\dd
    u^2+\frac{u^2}{4}(\dd \psi+P+P')^2\right)+\frac{u^2}{4F_2^2}\dd s^2(S^2)\right],
\eea   
with $H_{M5,M2}$, $F_{1,2}$ arbitrary functions of $u,t$, and $\a$ a
constant. To determine the Calabi-Yau interpolation with this ansatz,
we set $H_{M5,M2}=1$ and require that $F_{1,2}$ are functions only of
$u$. The derivation of the Calabi-Yau metric is now identical to that
for the three-fold interpolation of the previous subsection, as given
in \cite{me}.  This close analogy
between fivebranes wrapped on K\"{a}hler four-cycles in four-folds and
two-cycles in three-folds has recently been used to construct infinite
families of $AdS_3$ solutions \cite{toni1}, \cite{toni2}, \cite{dan}
motivated by 
the analogous $AdS_5$ solutions \cite{ypq}. 

In any event, to determine the special holonomy metric, observe that
closure of $\Omega_8$, with the obvious frame inherited from the $AdS$
solution, is automatic. Closure of $J_8$ results in the pair of
equations
\bea
\a^2\partial_u(F_1^2F_2^2)+\frac{u}{2F_1^2}&=&0,\nn
\partial_u\left(\frac{u^2}{4F_2^2}\right)-\frac{u}{2F_1^2}&=&0.
\eea
As in \cite{me}, \cite{us}, the general solution of these equations
inducing a metric with only one singular degeneration point is given
by
\bea
F_1^2&=&\frac{a^4}{\a^2u^2}\cos^2\xi,\nn
F_2^2&=&\frac{u^2}{2a^2}\frac{(1+\sin\xi)}{\cos^2\xi},\nn
-\frac{1}{3}\sin^2\xi+\sin\xi&=&\frac{2}{3}-\frac{\a^2u^4}{4a^6},
\eea
for some constant $\alpha$. Defining the coordinate 
\bea
R^2=\frac{\a^2 u^4}{4a^6},
\eea
the metric takes the form given above.

\begin{centering}
\subsection{Special Lagrangian Cycles}
\end{centering}

\noindent In this subsection we will give conjectured interpolating
pairs for fivebranes wrapped on SLAG cycles in three- and
four-folds. The $AdS$ solutions for which a Calabi-Yau interpolation
is derived are respectively the $AdS_4$ solution of \cite{sezgin},
admitting eight Killing spinors; and the $AdS_3$ solution of
\cite{gkw}, admitting four Killing spinors. In each case we will first
give the conjectured pair, then the derivation of the Calabi-Yau
interpolation from the $AdS$ solution.\\\\

\begin{centering}
\subsubsection{Three-fold}
\end{centering}

\paragraph{The interpolating pair} The eleven-dimensional lift of the
$AdS_4$ solution of \cite{sezgin} was later interpreted \cite{gkw} as
the near-horizon limit of fivebranes wrapped on a SLAG three-cycle in
a three-fold. The metric is given by 
\bea
\dd s^2=\frac{1}{\l}\left[\dd s^2(AdS_4)+\dd
  s^2(H^3)+(1-\l^3\r^2)\mbox{D}Y^a\mbox{D}Y^a+\frac{\l^3}{4(1-\l^3\r^2)}\Big(\dd
  \r^2+\r^2\dd s^2(S^1)\Big)\right],\nonumber
\eea
\bea\label{ads4}
\l^3&=&\frac{2}{8+\r^2}.
\eea
The flux, which in this case is purely magnetic and irrelevant to the
discussion, may be obtained from \cite{gkw} or \cite{wrap}. Here the $Y^a$, $a=1,2,3$, are constrained coordinates on an $S^2$,
$Y^aY^a=1$, and
\bea
\mbox{D}Y^a=\dd Y^a+\omega^a_{\;\;\;b}Y^b,
\eea
where the $\omega_{ab}$ are the spin-connection one-forms of
$H^3$. The range of $\rho$, which without loss of generality we take
to be positive, is $\rho\in[0,\sqrt{8}]$. At $\rho=0$ the R-symmetry
$S^1$ degenerates smoothly\footnote{The R-symmetry of the dual
  conformal theory is $U(1)$.}, while at $\rho=\sqrt{8}$ the $S^2$
degenerates smoothly. \\

\noindent Denoting a basis for $H^3$ by $e^a$, the metric of the
conjectured Calabi-Yau interpolation of this solution is  
\bea
\dd s^2=\dd s^2(\mathbb{R}^{1,4})+\dd s^2(\mathcal{N}_{\tau}),
\eea
where, up to an overall scale,
\bea
\dd
s^2(\mathcal{N}_{\tau})&=&\frac{(2\theta-\sin2\theta)^{1/3}}{\sin\theta}\Big[\frac{1}{2}(1-\cos\theta)(e^a-Y^aY^be^b)^2+\frac{1}{2}(1+\cos\theta)\mbox{D}Y^a\mbox{D}Y^a\nn&&+\frac{1}{3}\Big(\frac{\sin^3\theta}{2\theta-\sin2\theta}\Big)\Big(\dd\theta^2+4(Y^ae^a)^2\Big)\Big].
\eea
The range of $\theta$ is $\theta\in(0,\pi]$. Near $\theta=\pi$, the
$S^2$ degenerates smoothly; up to a scale, near $\theta=\pi$ the metric is
\bea
\dd s^2=\dd
s^2(H^3)+\frac{1}{4}\Big[\dd\theta^2+\theta^2\mbox{D}Y^a\mbox{D}Y^a\Big].
\eea
The metric is singular at $\theta=0$; up to a scale, near $\theta=0$ it is
\bea
\dd
s^2=\frac{1}{4}\Big[\dd\theta^2+\theta^2(e^a-Y^aY^be^b)^2\Big]+(Y^ae^a)^2+\mbox{D}Y^a\mbox{D}Y^a.
\eea
This Calabi-Yau is the hyperbolic analogue of the deformed conifold
\cite{cand} (which coincides with the Stenzel three-fold metric
\cite{stenzel}); the $S^3$ SLAG cycle of the deformed conifold is
replaced by a $H^3$ in the $\mathcal{N}_{\tau}$ metric. Now we discuss
the derivation of this interpolation from the $AdS$ solution.

\paragraph{The G-structure of the AdS solution} The $AdS_4$ solution
admits admits an $SU(2)$ structure defined by all eight Killing
spinors. It is given by \cite{wrap}
\bea
e^5&=&\frac{1}{\l^{1/2}}Y^ae^a,\nn
J^1&=&\frac{1}{\l}\sqrt{1-\l^3\r^2}\mbox{D}Y^a\wedge e^a,\nn
J^2&=&\frac{1}{\l}\sqrt{1-\l^3\r^2}\epsilon^{abc}Y^a\mbox{D}Y^b\wedge e^c,\nn
J^3&=&\frac{1}{2}\epsilon^{abc}\left[\frac{1}{\l}(1-\l^3\r^2)Y^a\mbox{D}Y^b\wedge \mbox{D}Y^c-\frac{1}{\l}Y^ae^b\wedge
e^c\right].
\eea 
This structure satisfies the torsion conditions of \cite{wrap} for the
near-horizon limit of fivebranes on a SLAG three-cycle in a three-fold, which
are
\bea
\dd\left(\l^{-1}\sqrt{1-\l^3\r^2}e^5\right)&=&
   \l^{-1/2}J^1+\l\r e^5\wedge\hat{\rho} ,\nn
\dd\left(\l^{-3/2}J^3\wedge e^5-\r J^2\wedge\hat{\rho}\right)
   &=&0,\nn
\dd\left(J^2\wedge e^5+\l^{-3/2}\r^{-1}J^3\wedge\hat{\rho}\right)
   &=&0.
\eea
The following identities, valid for a $H^3$ or $S^3$ with scalar
curvature $R$, are useful in verifying this claim:
\bea
\dd (Y^ae^a)&=&\mbox{D}Y^a\w e^a,\nn
\dd(\e^{abc}Y^a\mbox{D}Y^b\w\mbox{D}Y^c)&=&-\frac{R}{3}\e^{abc}Y^a\mbox{D}Y^b\w
e^c\w Y^de^d,\nn
\dd(\e^{abc}Y^a e^b\w e^c)&=&2\e^{abc}Y^a\mbox{D}Y^b\w
e^c\w Y^de^d,\nn\label{pl}
\dd(\e^{abc}Y^a\mbox{D}Y^b\w
e^c)&=&\e^{abc}\left[Y^a\mbox{D}Y^b\w\mbox{D}Y^c-\frac{R}{6}Y^ae^b\w
  e^c\right]\w Y^de^d.
\eea
In this case, the Bianchi identity for the flux is implied by the
torsion conditions \cite{wrap}.

\paragraph{The AdS solution in the Minkowski frame} From section 2, defining the
Minkowski-frame coordinates
\bea
t&=&-\frac{\r}{2}e^{-2r},\nn
u&=&-\sqrt{\frac{8-\r^2}{2}}e^{-r},
\eea
the metric of the $AdS$ solution in the Minkowski frame is given by
\bea
\dd s^2=L^{-1}\Big[\dd s^2(\mathbb{R}^{1,2})+F\dd
s^2(H^3)\Big]+L^2\Big[F^{-1}(\dd u^2+u^2\mbox{D}Y^a\mbox{D}Y^a)+\dd
s^2(\mathbb{R}^2)\Big],
\eea
where $L=\l e^{2r}$, $F=e^{2r}$ and 
\bea
e^{2r}=\frac{u^2}{4t^2}\left(-1+\sqrt{1+32t^2/u^2}\right).
\eea
The wrapped-brane $SU(3)$ structure of the $AdS$ solution, defined by
four of its Killing spinors, is given by
\bea
J_6&=&J^1+e^5\w \hat{u},\nn
\Omega_6&=&(J^3+iJ^2)\w(e^5+i\hat{u}),
\eea
with $\hat{u}=LF^{-1/2}\dd u$. By construction, this structure is a
solution of the wrapped brane equations for fivebranes wrapped on a
SLAG cycle in a three-fold, which are \cite{james}
\bea
\mbox{Vol}[\mathbb{R}^2]\w\dd\mbox{Im}\Omega_6&=&0,\nn
 \dd(L^{-1/2}J_6)&=&0,\nn
\mbox{Re}\Omega_6\w\dd\mbox{Re}\Omega_6&=&0,\nn
\dd \Big(\star_8L^{3/2}\dd(L^{-3/2}\mbox{Re}\Omega_6)\Big)&=&0,
\eea
where $\star_8$ denotes the Hodge dual on the space transverse to the
Minkowski factor.

\paragraph{The conjectured Calabi-Yau interpolation} We now make the
following ansatz for an interpolating solution:
\bea
\dd s^2&=&L^{-1}\Big[\dd
s^2(\mathbb{R}^{1,2})+F_1^2(e^a-Y^aY^be^b)^2+F_2^2(Y^ae^a)^2\Big]+L^2\Big[F_4^2\dd
u^2+F_3^2\mbox{D}Y^a\mbox{D}Y^a+\dd s^2(\mathbb{R}^2)\Big],\nn&&
\eea
with $L$, $F_{1,...,4}$ arbitrary functions of $u$ and $t$. To
determine the Calabi-Yau interpolation with this ansatz, we set $L=1$,
and require that $F_{1,...,4}$ are functions only of $u$. Then $F_4$
is at our disposal and we set it to unity. The Calabi-Yau condition is
\bea
\dd J_6=\dd\Omega_6=0,
\eea
with $J_6$ and $\Omega_6$ as inherited from the $AdS$ solution in the
Minkowski frame,
\bea
J_6&=&F_1F_3\mbox{D}Y^a\w e^a+F_2Y^a e^a\w\dd u,\nn
\mbox{Re}\Omega_6&=&\frac{1}{2}\Big(F_2F_3^2\e^{abc}Y^a\mbox{D}Y^b\w\mbox{D}Y^c-F_1^2F_2\e^{abc}Y^a
  e^b\w e^c\Big)\w Y^de^d-F_1F_3\e^{abc}Y^a\mbox{D}Y^b\w e^c,\nn
\mbox{Im}\Omega_6&=&F_1F_2F_3\e^{abc}Y^a\mbox{D}Y^b\w e^c\w
Y^de^d+\frac{1}{2}\Big(F_3^2\e^{abc}Y^a\mbox{D}Y^b\w\mbox{D}Y^c-F_1^2\e^{abc}Y^ae^b\w
e^c\Big)\w\dd u.\nn&&
\eea
Then using the equations \eqref{pl}, closure of $J_6$ implies
\bea\label{kj}
\partial_u(F_1F_3)+F_2=0.
\eea
Closure of $\mbox{Re}\Omega_6$ implies
\bea
\frac{1}{2}\partial_u(F_2F_3^2)+F_1F_3&=&0,\nn\label{kjh}
\frac{1}{2}\partial_u(F_2F_1^2)-F_1F_3&=&0.
\eea
Closure of $\mbox{Im}\Omega_6$ implies
\bea
\partial_u(F_1F_2F_3)-F_3^2+F_1^2=0,
\eea
and this equation is implied by the other three. Solving \eqref{kj}
and \eqref{kjh} is straightforward. Adding \eqref{kjh} we immediately
get
\bea
F_2=\frac{a}{F_1^2+F_3^2},
\eea
for constant $a$. Next, subtracting \eqref{kjh}, and defining a new
coordinate $x$ according to
\bea
\partial_u=-\frac{4}{a}F_1F_3\partial_x,
\eea
we get
\bea
\frac{F_3^2-F_1^2}{F_1^2+F_3^2}=x+b,
\eea
for a constant $b$ which may be eliminated by a shift of $x$. Solving
for $F_3$, inserting in \eqref{kj}, and defining $x=\cos\theta$, we
obtain
\bea
F_1^6=\frac{3a^2}{32}(2\theta-\sin2\theta+c)\left(\frac{1-\cos\theta}{1+\cos\theta}\right)^{3/2},
\eea
for constant $c$. The metric has pathological behaviour unless $c=0$,
so we choose this value. Then, up to an overall scale of
$(3a^2/4)^{1/3}$, we obtain the three-fold metric given above. 

\begin{centering}
\subsubsection{Four-folds}
\end{centering}

\paragraph{The interpolating pair} The GKW solution for $AdS_3$ near-horizon limit of
a string intersection of fivebranes wrapped on a SLAG four-cycle in a
four-fold, with
membranes extended in the directions transverse to the four-fold, was
constructed in \cite{gkw}. The metric is given by
\bea
\dd s^2=\frac{1}{\l}\left[\dd s^2(AdS_3)+\frac{8}{3}\dd
  s^2(H^4)+(1-\l^3f^2)\mbox{D}Y^a\mbox{D}Y^a+\frac{\l^3}{4(1-\l^3f^2)}\dd\r^2\right],\nonumber
\eea
\bea
\l^3=\frac{16}{24+3\r^2},\;\;f=\frac{3\r}{4}.
\eea   
Here the $Y^a$, $a=1,...,4$ are constrained coordinates on a
three-sphere, $Y^aY^a=1$, and 
\bea
\mbox{D}Y^a=\dd Y^a+\omega^a_{\;\;\;b}Y^b,
\eea
with $\omega_{ab}$ the spin connection one-forms of $H^4$. The range
of $\rho$ is $\rho\in[-2,2]$; at the endpoints, the $S^3$ degenerates
smoothly. The electric flux may be obtained from \cite{gkw} or
\cite{pau}; the magnetic flux will be given below.\\

\noindent Denoting a basis for $H^4$ by $e^a$, the metric of the
conjectured Calabi-Yau interpolation of this solution is
\bea
\dd s^2=\dd s^2(\mathbb{R}^{1,2})+\dd s^2(\mathcal{N}_{\tau}),
\eea
where, up to an overall scale, 
\bea
\dd
s^2(\mathcal{N}_{\tau})&=&\frac{(2+\cos2\theta)^{1/4}}{\cos\theta}\Big[\cos^2\theta(e^a-Y^aY^be^b)^2+\sin^2\theta\mbox{D}Y^a\mbox{D}Y^a\nn&&+\frac{3\cos\theta\sin^32\theta}{8\sin^3\theta(2+\cos2\theta)}\Big(\dd\theta^2+(Y^ae^a)^2\Big)\Big].
\eea
This metric is the hyperbolic analogue of the Stenzel four-fold. Without loss of generality, we can take the range of $\theta$ to be
$\theta\in[0,\pi/2)$. Near $\theta=0$, the $S^3$ degenerates smoothly,
and up to a scale the metric is given by
\bea
\dd s^2=\dd s^2(H^4)+\dd\theta^2+\theta^2\mbox{D}Y^a\mbox{D}Y^a.
\eea
The other degeneration point, $\theta=\pi/2$, is singular. Now we give
the derivation of the conjectured interpolation.

\paragraph{The G-structure of the AdS solution} The $AdS_3$ solution
admits an $SU(3)$ structure defined by all four Killing spinors. The
structure satisfies the torsion
conditions of \cite{pau} for the near-horizon limit of fivebranes on a
SLAG four-cycle in a four-fold, together with the Bianchi identity for the
magnetic flux, $\dd F_{\mbox{\small{mag}}}=0$, which in this case is
not implied by the torsion conditions. The $SU(3)$ structure is
\cite{pau} 
\bea
e^7&=&-\sqrt{\frac{8}{3\l}}Y^ae^a,\nn
J_6&=&\sqrt{\frac{8(1-\l^3 f^2)}{3\l^2}}e^a\w \mbox{D}Y^a,\nn
\mbox{Re}\Omega_6&=&\left(\sqrt{\frac{8}{3\l}}\right)^3\frac{1}{3!}\e^{abcd}Y^ae^b\w e^c\w
e^d-\sqrt{\frac{8}{3\l^3}}\Big(1-\l^3 f^2\Big)\frac{1}{2}\e^{abcd}Y^a\mbox{D}Y^b\w \mbox{D}Y^c\w e^d,\nn
\mbox{Im}\Omega_6&=&\frac{8}{3}\sqrt{\frac{1-\l^3f^2}{\l^3}}\frac{1}{2}\e^{abcd}Y^a\mbox{D}Y^b\w e^c\w
e^d-\left(\sqrt{\frac{1-\l^3f^2}{\l}}\right)^3\frac{1}{3!}\e^{abcd}Y^a\mbox{D}Y^b\w\mbox{D}Y^c\w\mbox{D}Y^d.\nn&&
\eea
The torsion conditions are
\bea
e^7\w\hat{\r}\w\dd\left(\frac{\mbox{Re}\Omega_6}{\sqrt{1-\l^3f^2}}\right)&=&0, \label{eqn:slag4Re}\\
\dd\left(\l^{-1}\sqrt{1-\l^3f^2} e^7\right)&=&\l^{-1/2}(J_6+\l^{3/2}f e^7\w\hat{\r}),\label{eqn:slag4de7}\\
\mbox{Im}\Omega_6\w\dd\mbox{Im}\Omega_6&=&\frac{\l^{1/2}}{\sqrt{1-\l^3f^2}}(6+4\l^3f^2)\mbox{Vol}[\mathcal{M}_6]\w
e^7-2\l^{3/2}f\star_8\dd\log\left(\frac{\l^3f}{1-\l^3f^2}\right),\nn
\label{eqn:slag4Im}
\eea
where 
\bea
\mbox{Vol}[\mathcal{M}_6]=\frac{1}{3!}J_6\w J_6\w J_6.
\eea
and $\star_8$ denotes the Hodge dual on the space transverse to the
$AdS$ factor, with positive orientation defined with respect to
\bea
\mbox{Vol}=\mbox{Vol}[\mathcal{M}_6]\w e^7\w \hat{\r}.
\eea 
The magnetic flux is
\bea
F_{\mbox{\small{mag}}}&=&-\;\frac{\l^{3/2}}{1-\l^3f^2}(\l^{3/2}f+\star_8)\left(\dd\left[\l^{-3/2}\sqrt{1-\l^3f^2}\mbox{Im}\Omega_6\right]+4\l^{-1}\mbox{Re}\Omega_6\w
e^7\right)-2\l^{1/2}\mbox{Im}\Omega_6\w\hat{\r}.\nn
\eea
The following identities, valid for a $H^4$ or an $S^4$ with scalar
curvature $R$, are useful in verifying the torsion conditions and
Bianchi identity:
\bea
\dd\Big(\e^{abcd}Y^ae^b\w e^c\w e^d\Big)&=&-3\e^{abcd}Y^a\mbox{D}Y^b\w
e^c\w e^d\w Y^ee^e,\nn
\dd\Big(\e^{abcd}Y^a\mbox{D}Y^b\w e^c\w
e^d\Big)&=&\Big(-2\e^{abcd}Y^a\mbox{D}Y^b\w \mbox{D}Y^c\w
e^d+\frac{R}{12}\e^{abcd}Y^ae^b\w e^c\w e^d\Big)\w Y^ee^e,\nn
\dd\Big(\e^{abcd}Y^a\mbox{D}Y^b\w \mbox{D}Y^c\w
e^d\Big)&=&\Big(\frac{R}{6}\e^{abcd}Y^a\mbox{D}Y^b\w e^c\w
e^d-\e^{abcd}Y^a\mbox{D}Y^b\w\mbox{D}Y^c\w\mbox{D}Y^d\Big)\w
Y^ee^e,\nn\label{su4id} 
\dd\Big(\e^{abcd}Y^a\mbox{D}Y^b\w\mbox{D}Y^c\w\mbox{D}Y^d\Big)&=&\frac{R}{4}\e^{abcd}Y^a\mbox{D}Y^b\w
\mbox{D}Y^c\w e^d\w Y^ee^e.
\eea

\paragraph{The AdS solutions in the Minkowski frame} Using section 2,
we define the coordinates
\bea
t&=&-\frac{1}{2}e^{-3r/2}\r,\nn
u&=&-\sqrt{\frac{24-6\r^2}{16}}e^{-r},
\eea
so that the one-forms $e^8$, $e^9$ in the Minkowski frame are given by
\bea
e^8&=&\l e^r\dd u,\nn
e^9&=&\l e^{3r/2}\dd t,
\eea
and the $AdS$ metric in the Minkowski frame takes the form
\bea
\dd s^2&=&\frac{1}{H^{1/3}_{M5}H_{M2}^{2/3}}\dd
s^2(\mathbb{R}^{1,1})+\frac{H^{2/3}_{M5}}{H^{2/3}_{M2}}\dd
t^2+\frac{H^{1/3}_{M2}}{H^{1/3}_{M5}}\left[\frac{8}{3}F\dd
  s^2(H^4)\right]\nn&&+H^{1/3}_{M2}H_{M5}^{2/3}\left[\frac{1}{F}\left(\dd
    u^2+u^2\mbox{D}Y^a\mbox{D}Y^a\right)\right],
\eea
where
\bea
H_{M5}&=&\l^3 e^{5r},\nn
H_{M2}&=&e^{r/2},\nn
F&=&e^{3r/2}.
\eea 
The function $e^{r}$ is given in terms of $t$ and $u$ by a positive
signature metric inducing root of the cubic
\bea
t^{2}e^{3r}+\frac{2}{3}u^2e^{2r}-1=0.
\eea
The wrapped brane $SU(4)$ structure of the $AdS_3$ solution, defined
by two of its Killing spinors, is
given by
\bea
J_8&=&J_6+e^7\w e^8,\nn
\Omega_8&=&\Omega_6\w(e^7+ie^8).
\eea
By construction, this structure is a solution of the wrapped brane
equations for a SLAG four-cycle in a four-fold, which
comprise the torsion conditions \cite{pau}
\bea
\dd(L^{-1/2}J_8)&=&0,\nn
\mbox{Im}\Omega_8\w \dd\mbox{Re}\Omega_8&=&0,\nn
e^9\w[\mbox{Re}\Omega_8\lrcorner \;\dd\mbox{Re}\Omega_8-2L^{3/2}e^9\lrcorner\;\dd(L^{-3/2}e^9)]&=&0.
\eea
together with the Bianchi identity and field equation for the
four-form, which is given in \cite{pau}.

\paragraph{The conjectured Calabi-Yau interpolation} We make the
following ansatz for an interpolating solution:
\bea
\dd s^2&=&\frac{1}{H^{1/3}_{M5}H_{M2}^{2/3}}\dd
s^2(\mathbb{R}^{1,1})+\frac{H^{2/3}_{M5}}{H^{2/3}_{M2}}\dd
t^2+\frac{H^{1/3}_{M2}}{H^{1/3}_{M5}}\left[F_1^2(e^a-Y^aY^be^b)^2+F_2^2(Y^be^b)^2\right]\nn&&+H^{1/3}_{M2}H_{M5}^{2/3}\left[F_4^2\dd
    u^2+F_3^2\mbox{D}Y^a\mbox{D}Y^a\right],
\eea   
with $H_{M5,M2}$, $F_{1,...,4}$ arbitrary functions of $u,t$. To
determine the Calabi-Yau interpolation with this ansatz,
we set $H_{M5,M2}=1$ and require that $F_{1,...,4}$ are functions only of
$u$. Then $F_4$ is at our disposal, and we set it to 1. Requiring
$SU(4)$ holonomy, we set
\bea
\dd J_8=\dd\Omega_8=0,
\eea
with
\bea
J_8&=&J_6+e^7\w\dd u,\nn
\Omega_8&=&\Omega_6\w(e^7+i\dd u),\nn
e^7&=&-F_2Y^ae^a,\nn
J_6&=&F_1F_3e^a\w\mbox{D}Y^a,\nn
\mbox{Re}\Omega_6&=&F_1^3\frac{1}{3!}\e^{abcd}Y^ae^b\w e^c\w
e^d-F_1F_3^2\frac{1}{2}\e^{abcd}Y^a\mbox{D}Y^b\w \mbox{D}Y^c\w e^d,\nn
\mbox{Im}\Omega_6&=&F_1^2F_3\frac{1}{2}\e^{abcd}Y^a\mbox{D}Y^b\w e^c\w
e^d-F_3^3\frac{1}{3!}\e^{abcd}Y^a\mbox{D}Y^b\w\mbox{D}Y^c\w\mbox{D}Y^d.
\eea 
Closure of $J_8$ implies
\bea
\partial_u(F_1F_3)+F_2=0.
\eea
Using the identities \eqref{su4id} with $R=-12$, closure of
$\mbox{Re}\Omega_8$ implies
\bea
\partial_u(F_1^3F_2)-3F_1^2F_3&=&0,\nn
\partial_u(F_3^2F_2)+3F_1F_3^2&=&0,
\eea
while closure of $\mbox{Im}\Omega_8$ implies
\bea
\partial_u(F_1^2F_2F_3)+F_1^3-2F_1F_3^2&=&0,\nn
\partial_u(F_1F_2F_3^2)-F_3^3+2F_1^2F_3&=&0.
\eea
It may be verified that the last two equations are implied by the
first three. Solving for $F_{1,2,3}$ is straightforward. First define
a new coordinate $x$ according to
\bea
-F_2\partial_u=\partial_{\theta}.
\eea
Then we have that
\bea
\partial_{\theta}\left(\frac{F_1}{F_3}\right)=-1-\left(\frac{F_1}{F_3}\right)^2,
\eea
which has solution
\bea
\frac{F_1}{F_3}=\frac{\cos\theta}{\sin\theta},
\eea
up to an irrelevant constant which may be eliminated by a shift of
$\theta$. Using this, we find that
\bea
\partial_{\theta}\log\left(\frac{F_1F_2^{2/3}F_3}{\sin 2\theta}\right)=0,
\eea
and hence that
\bea
F_1F_2^{2/3}F_3=\a\sin2\theta,
\eea
for constant $\a$. Finaly we get
\bea
\partial_{\theta}\left(\frac{\sin2\theta}{F_2^{2/3}}\right)=\frac{1}{\a}F_2^2,
\eea
which has solution
\bea
F_2=\left(\frac{3\a}{8}\right)^{3/8}\left[\frac{\sin2\theta}{(\beta+[2+\cos2\theta]\sin^4\theta)^{1/4}}\right]^{3/2},
\eea
for constant $\beta$. As was the case for the three-fold solution of
the previous subsection, the metric has pathological behaviour unless
$\beta=0$. Choosing this value, the metric, up to an overall scale of
$(8\a^3/3)^{1/4}$, is as given above.

\begin{centering}
\section{Sp(2) interpolating pair}
\end{centering}

\noindent In this section, we will give a conjectured interpolating
pair for fivebranes wrapped on a complex lagarangian four-cycle in an $Sp(2)$
manifold. First we give the pair, then the derivation of the $Sp(2)$
interpolation from the $AdS$ solution.\\

\paragraph{The interpolating pair} In \cite{jk}, an $AdS_3$ solution
admitting six Killing spinors and  
describing the near-horizon limit of fivebranes wrapped on a CLAG
four-cycle in an $Sp(2)$ manifold was constructed. In addition to
the fivebranes, there are membranes extended in the directions
transverse to the $Sp(2)$, which intersect the fivebranes in a
string. The quantum dual of the $AdS$ solution is the two-dimensional
low energy effective theory on the string worldvolume. The metric of
the $AdS$ solution is given by
\bea
\dd s^2=\frac{1}{\l}\left[\dd s^2(AdS_3)+\frac{5}{2}\dd
  s^2(\mbox{B}_4)+(1-\l^3f^2)\mbox{D}Y^a\mbox{D}Y^a+\frac{\l^3}{4(1-\l^3f^2)}\dd\r^2\right],\nonumber
\eea
\bea\label{sp2}
\l^3=\frac{50}{60+3\r^2},\;\;f=\frac{3\r}{5}.
\eea 
Here $\dd s^2(\mbox{B}_4)$ is the Bergman metric on two-dimensional
complex hyperbolic space, normalised such that the scalar curvature is
$R=-12$; explicitly, this metric is
\bea\label{Bergman}
\dd s^2(\mbox{B}_4)=2\left[\dd
  z^2+\frac{1}{4}\sinh^2z(\s_1^2+\s_2^2+\cosh^2z\s_3^2)\right],
\eea
with $\dd\s_1=\s_2\w\s_3$, together with cyclic permutations. In the
$AdS$ metric \eqref{sp2}, the $Y^a$, $a=1,...,4$ parameterise an
$S^3$, $Y^aY^a=1$, and
\bea
\mbox{D}Y^a=\dd Y^a+\omega^a_{\;\;\;b}Y^b,
\eea
with $\omega_{ab}$ the spin connection one-forms of $\mbox{B}_4$. The
electric flux is irrelevant to the discussion, and may be obtained
from \cite{jk} or \cite{pau}; the magnetic flux will be given
below.\\

\noindent To give the conjectured special holonomy interpolation of
this metric, we first make the following definitions. Let $e^a$ denote
a basis for the Bergman metric \eqref{Bergman}. Let $J^A$, $A=1,2,3$,
denote a basis of self-dual $SU(2)$ invariant three-forms on
$\mbox{B}_4$, obeying the algebra $J^AJ^B=-\delta^{AB}-\e^{ABC}J^C$,
and let $J^3$ be the K\"{a}hler form of $\mbox{B}_4$. Define 
\bea
\mbox{E}_1=J^1_{ab}Y^ae^b,\;\;\mbox{E}_2=-J^2_{ab}Y^ae^b,\;\;\mbox{E}_3=J^1_{ab}Y^a\mbox{D}Y^b,\;\;\mbox{E}_4=J^2_{ab}Y^a\mbox{D}Y^b,\nonumber
\eea
\bea
\mbox{E}_5=J^3_{ab}Y^ae^b,\;\;
\mbox{E}_6=Y^ae^a,\;\;\mbox{E}_7=J^3_{ab}Y^a\mbox{D}Y^b.
\eea
Then the conjectured hyper-K\"{a}hler interpolation of the $AdS_3$
solution is
\bea
\dd s^2=\dd s^2(\mathbb{R}^{1,2})+\dd s^2(\mathcal{N}_{\tau}),
\eea
where, up to an overall scale,
\bea
\dd
s^2(\mathcal{N}_{\tau})&=&\Big(1+R^2\Big)\Big(\mbox{E}_1^2+\mbox{E}_2^2\Big)+2\Big(1-R^2\Big)\Big(\mbox{E}_3^2+\mbox{E}_4^2\Big)+2R^2\Big(\mbox{E}_5^2+\mbox{E}_6^2\Big)\nn&&+R^2\left(\frac{1}{R^4}-1\right)\mbox{E}_7^2+4\left(\frac{1}{R^4}-1\right)^{-1}\dd
R^2.
\eea
The range of $R$ is $R\in(0,1]$. At $R=1$, the $S^3$ degenerates
smoothly. Defining $R=1-y/2$, the metric near $y=0$ is
\bea
\dd s^2(\mathcal{N}_{\tau})=2\dd s^2(\mbox{B}_4)+\dd
y^2+y^2\mbox{D}Y^a\mbox{D}Y^a.
\eea
The metric is singular at $R=0$. This $\mathcal{N}_{\tau}$ metric is the hyperbolic
analogue of the Calabi metric on $T^*\mathbb{CP}^2$ \cite{calabi}. Now
we give its derivation from the $AdS$ solution.

\paragraph{The G-structure of the AdS solution} The $AdS_3$ admits an
$SU(2)$ structure defined by all six Killing spinors. This structure
satisfies the torsion conditions of \cite{pau}, for the near-horizon
limit of fivebranes on a CLAG four-cycle, together with the Bianchi
identity for the flux $\dd F_{\mbox{\small{mag}}}=0$, which in this case is not implied by the
torsion conditions. The $SU(2)$ structure is given by
\bea
e^5=\sqrt{\frac{5}{2\l}}\mbox{E}_5,\;\;e^6&=&\sqrt{\frac{5}{2\l}}\mbox{E}_6,\;\;e^7=\sqrt{\frac{1-\l^3f^2}{\l}}E_7,\nonumber
\eea
\bea
K^1&=&\frac{1}{\l}\sqrt{\frac{5(1-\l^3f^2)}{2}}\Big(\mbox{E}_1\w\mbox{E}_4+\mbox{E}_2\w\mbox{E}_3\Big),\nn
K^2&=&\frac{1}{\l}\sqrt{\frac{5(1-\l^3f^2)}{2}}\Big(-\mbox{E}_1\w\mbox{E}_3+\mbox{E}_2\w\mbox{E}_4\Big),\nn
K^3&=&\frac{5}{2\l}\mbox{E}_1\w\mbox{E}_2+\frac{(1-\l^3f^2)}{\l}\mbox{E}_3\w\mbox{E}_4.
\eea
The triplet of $SU(2)$ structure forms $K^A$ (not to be confused with
the $J^A$ forms on $\mbox{B}_4$) obey the algebra $K^AK^B=-\delta^{AB}-\e^{ABC}K^C$.
The relevant torsion conditions of \cite{pau} are 
\bea
\hat{\r}\w\dd\Big[\l^{-1}\left(\mbox{Vol}[\mathcal{M}_4]+K^3\w e^{56}\right)\Big]&=&0,\\
(K^3+e^{56})\lrcorner\;\dd
e^7&=&\frac{2\l^{1/2}}{\sqrt{1-\l^3 f^2}}(1+\l^3f^2)-\l^{3/2}f\hat{\r}\lrcorner\;\dd\log\left(\frac{\l^3f}{1-\l^3f^2}\right),\nn
\dd\left(\l^{-1}\sqrt{1-\l^3f^2}e^5\right)&=&\l^{-1/2}\Big(K^1+e^{67}+\l^{3/2}f e^5\w\hat{\r}\Big),\\
\dd\left(\l^{-1}\sqrt{1-\l^3f^2} e^6\right)&=&\l^{-1/2}\Big(K^2+e^{75}+\l^{3/2}f e^2\w\hat{\r}\Big),
\eea
with
\bea
\mbox{Vol}[\mathcal{M}_4]=\frac{1}{2}K^3\w K^3.
\eea
The magnetic flux is
\bea
F_{\mbox{\small{mag}}}&=&\frac{\l^{3/2}}{1-\l^3f^2}(\l^{3/2}f+\star_8)\Big[\dd\left(\l^{-3/2}\sqrt{1-\l^3f^2}\Big[K^3\w
e^7+e^{567}\Big]\right)\nn&&-4\l^{-1}\Big(\mbox{Vol}[\mathcal{M}_4]+K^3\w
e^{56}\Big)\Big]+2\l^{1/2}\Big(K^3\w e^7+e^{567}\Big)\w\hat{\r},
\eea 
with
\bea
\mbox{Vol}[\mathcal{M}_8]=\mbox{Vol}[\mathcal{M}_4]\w
e^{567}\w\hat{\r}.
\eea
In verifying that the given structure indeed solves the torsion
conditions and Bianchi identity, and in the derivation of the $Sp(2)$ metric
to follow, the following is useful. Defining
\bea
Q=\frac{1}{2}J^{3ab}\omega_{ab},
\eea
the exterior derivatives of the Es are given by
\bea
\dd\mbox{E}_1&=&-\mbox{E}_2\w(Q+\mbox{E}_7)-\mbox{E}_3\w\mbox{E}_6+\mbox{E}_4\w\mbox{E}_5,\nn
\dd\mbox{E}_2&=&\mbox{E}^1\w(Q+\mbox{E}_7)+\mbox{E}_3\w\mbox{E}_5+\mbox{E}_4\w\mbox{E}_6,\nn
\dd\mbox{E}_3&=&\mbox{E}_4\w(Q+2\mbox{E}_7)-\frac{1}{2}\mbox{E}_1\w\mbox{E}_6+\frac{1}{2}\mbox{E}_2\w\mbox{E}_5,\nn
\dd\mbox{E}_4&=&-\mbox{E}_3\w(Q+2\mbox{E}_7)+\frac{1}{2}\mbox{E}_2\w\mbox{E}_6+\frac{1}{2}\mbox{E}_1\w\mbox{E}_5,\nn
\dd\mbox{E}_5&=&\mbox{E}_1\w\mbox{E}_4+\mbox{E}_2\w\mbox{E}_3+\mbox{E}_6\w\mbox{E}_7,\nn
\dd\mbox{E}_6&=&-\mbox{E}_1\w\mbox{E}_3+\mbox{E}_2\w\mbox{E}_4+\mbox{E}_7\w\mbox{E}_5,\nn
\dd\mbox{E}_7&=&-\mbox{E}_1\w\mbox{E}_2+2\mbox{E}_3\w\mbox{E}_4-2\mbox{E}_5\w\mbox{E}_6.
\eea

\paragraph{The AdS solution in the Minkowski frame} We now use section
2 to frame-rotate the $AdS$ solution. Defining the coordinates
\bea
t&=&-\frac{1}{2}e^{-6r/5}\r,\nn
u&=&-\sqrt{\frac{12-3\r^2}{10}}e^{-r},
\eea
the one-forms $e^8$, $e^9$ in the Minkowski frame are given by
\bea
e^8&=&\l e^r\dd u,\nn
e^9&=&\l e^{6r/5}\dd t,
\eea
and the $AdS$ metric in the Minkowski frame takes the form
\bea
\dd s^2&=&\frac{1}{H^{1/3}_{M5}H_{M2}^{2/3}}\dd
s^2(\mathbb{R}^{1,1})+\frac{H^{2/3}_{M5}}{H^{2/3}_{M2}}\dd
t^2+\frac{H^{1/3}_{M2}}{H^{1/3}_{M5}}\left[\frac{5}{2}F\dd
  s^2(\mbox{B}_4)\right]\nn&&+H^{1/3}_{M2}H_{M5}^{2/3}\left[\frac{1}{F}\left(\dd
    u^2+u^2\mbox{D}Y^a\mbox{D}Y^a\right)\right],
\eea
where
\bea
H_{M5}&=&\l^3 e^{22r/5},\nn
H_{M2}&=&e^{4r/5},\nn
F&=&e^{6r/5}.
\eea 
The function $e^{2r}$ is given in terms of $t$ and $u$ by a positive
signature metric inducing root of the sextic
\bea
t^{2}e^{12r}-\left(1-\frac{5}{6}u^2e^{2r}\right)^5=0.
\eea
The wrapped brane $Sp(2)$ structure of the $AdS_3$ solution, defined
by three of its Killing spinors, is
given by
\bea
J^1&=&K^3+e^5\w e^6+e^7\w e^8,\nn
J^2&=&K^2-e^5\w e^7+e^6\w e^8,\nn
J^3&=&K^1+e^6\w e^7+e^5\w e^8.
\eea
By construction, this structure is a solution of the wrapped brane
equations for a CLAG four-cycle in a hyper-K\"{a}hler eight-manifold, which
comprise the torsion conditions \cite{pau}
\bea
\dd(L^{-1/2}J^2)=\dd(L^{-1/2}J^3)&=&0,\nn
e^9\w[J^1\lrcorner \;\dd J^1-Le^9\lrcorner\;\dd(L^{-1}e^9)]&=&0,
\eea
together with the Bianchi identity and field equation for the
four-form, which is given in \cite{pau}.

\paragraph{The conjectured hyper-K\"{a}hler interpolation} We make the
following ansatz for an interpolating solution:
\bea
\dd s^2&=&\frac{1}{H^{1/3}_{M5}H_{M2}^{2/3}}\dd
s^2(\mathbb{R}^{1,1})+\frac{H^{2/3}_{M5}}{H^{2/3}_{M2}}\dd
t^2+\frac{H^{1/3}_{M2}}{H^{1/3}_{M5}}\left[F_1^2\Big(\mbox{E}_1^2+\mbox{E}_2^2\Big)+F_2^2\Big(\mbox{E}_5^2+\mbox{E}_6^2\Big)\right]\nn&&+H^{1/3}_{M2}H_{M5}^{2/3}\left[F_5^2\dd
    u^2+F_3^2\Big(\mbox{E}_3^2+\mbox{E}_4^2\Big)+F_4^2\mbox{E}_7^2\right],
\eea   
with $H_{M5,M2}$, $F_{1,...,5}$ arbitrary functions of $u,t$. To
determine the hyper-K\"{a}hler interpolation with this ansatz,
we set $H_{M5,M2}=1$ and require that $F_{1,...,5}$ are functions only of
$u$. Then $F_5$ is at our disposal, and we set it to 1. Requiring
$Sp(2)$ holonomy, we set
\bea
\dd J^A=0.
\eea
From $\dd J^1$, we derive the conditions
\bea
\partial_u\left(F_1^2\right)&=&F_4,\nn
\partial_u\left(F_2^2\right)&=&2F_4,\nn
\partial_u\left(F_3^2\right)&=&-2F_4,\nn
F_1^2&=&F_2^2+\frac{1}{2}F_3^2.
\eea
The algebraic constraint, combined with any two of the differential
equations, implies the third. From $\dd J^2$, we get
\bea
\partial_u\left(F_1F_3\right)&=&-F_2,\nn
\partial_u\left(F_2F_4\right)&=&-F_2,\nn\label{r}
F_1F_3&=&F_2F_4,
\eea
and from $\dd J^3$ we again obtain the equations \eqref{r}. The
algebraic constraint in \eqref{r}, combined with either of the
differential equations, implies the second. Therefore the system we
need to solve is
\bea
\partial_u\left(F_1^2\right)&=&F_4,\nn
\partial_u\left(F_2^2\right)&=&2F_4,\nn
\partial_u\left(F_2F_4\right)&=&-F_2,\nn
F_3^2&=&2\left(F_1^2-F_2^2\right),\nn\label{y}
F_1F_3&=&F_2F_4.
\eea
To solve the system, define a new coordinate $x$ such that
\bea
\partial_u=F_4\partial_x.
\eea
Then the first two equations of \eqref{y} give
\bea
F_1^2&=&x+a,\nn
F_2^2&=&2x+b,
\eea
for constants $a$, $b$. We eliminate $b$ by a shift of
$x$. Integrating the third equation we get
\bea
F_4^2=\frac{c}{x}-x,
\eea
for a constant $c$. Then the algebraic conditions imply that
\bea
F_3^2&=&2(a-x),\nn
c&=&a^2.
\eea
Finally, defining a new coordinate $x=aR^2$, up to an overall scale of
$a$, we get the hyper-K\"{a}hler $\mathcal{N}_{\tau}$ metric given above.

\begin{centering}
\section{$G_2$ interpolating pairs}
\end{centering}

\noindent In this section, we will give conjectured interpolating
pairs for fivebranes wrapped on calibrated cycles in $G_2$
manifolds. First we will discuss co-associative four-cycles, then
associative three-cycles. In each case we will first give the
conjectured pairs, followed by the derivation of the $G_2$
interpolations from the $AdS$ solutions.\\\\
 
\begin{centering}
\subsection{Co-associative cycles}
\end{centering}

\paragraph{The interpolating pairs} The GKW $AdS_3$ solutions \cite{gkw}, describing the near-horizon limit
of M-fivebranes wrapped on a co-associative cycle in a manifold of
$G_2$ holonomy, admit four Killing spinors, and have metrics 
\bea
\dd s^2=\frac{1}{\l}\left[\dd s^2(AdS_3)+\frac{9}{4}\dd
  s^2(\Sigma_4)+\frac{9}{4}(1-\l^3\r^2)\mbox{D}Y^a\mbox{D}Y^a+\frac{\l^3}{4}\left(\frac{\dd\r^2}{1-\l^3\r^2}+\r^2\dd
    s^2(S^1)\right)\right],\nonumber
\eea
\bea
\l^3=\frac{81}{64+54\r^2}.
\eea
In this case the flux is purely magnetic, and is irrelevant to the
discussion; it may be obtained from \cite{gkw} or \cite{wrap}. The
wrapped cycle $\Sigma_4$ is an arbitrary conformally half-flat Einstein manifold, with
scalar curvature normalised such that $R=-12$. This means that the
Ricci tensor of $\Sigma_4$ is given by 
\bea
R_{ij}=-3g_{ij},
\eea
and the Weyl tensor is anti-self-dual,
\bea
J^{aij}_4C_{ijkl}=0,
\eea
for a triplet of self-dual two-forms $J^a_4$, $a=1,2,3$, on $\Sigma_4$. An example
of such a manifold is hyperbolic four-space $H^4$. The
$Y^a$ are constrained coordinates on $S^2$, $Y^aY^a=1$, and
\bea
\mbox{D}Y^a=\dd Y^a-\frac{1}{2}\e^{abc}Y^b\omega_{ij}J_4^{cij},
\eea
where $\omega_{ij}$ are the spin connection one-forms of
$\Sigma_4$. The range of $\rho$, which without loss of generality is
taken to be non-negative, is $\rho\in[0,8/3\sqrt{3}]$. At $\rho=0$ the
R-symmetry $S^1$ degenerates smoothly\footnote{The R-symmetry of the
  conformal duals is $U(1)$.}, while at $\r=8/3\sqrt{3}$ the
$S^2$ parameterised by the $Y^a$ degenerates smoothly.\\ 

\noindent The metric of the conjectured $G_2$ interpolation of these
$AdS$ solutions is
\bea
\dd s^2=\dd s^2(\mathbb{R}^{1,3})+\dd s^2(\mathcal{N}_{\tau}), 
\eea
where up to an overall scale,
\bea
\dd s^2(\mathcal{N}_{\tau})=\frac{R^2}{2}\dd
s^2(\Sigma_4)+\frac{R^2}{4}\left(\frac{1}{R^4}-1\right)\mbox{D}Y^a\mbox{D}Y^a+\left(\frac{1}{R^4}-1\right)^{-1}\dd
R^2.
\eea
The range of $R$ is $R\in(0,1]$. At $R=1$, the $S^2$ degenerates
smoothly. The metric is singular at $R=0$ where the co-associative
$\Sigma_4$ degenerates. These metrics are the analogues, for
negatively curved conformally 
half-flat Einstein $\Sigma_4$, of the regular BSGPP $G_2$ metrics
on $\mathbb{R}^3$ bundles over $S^4$ or $\mathbb{CP}^2$ \cite{bryant},
\cite{pope}. Now we give their derivation from the $AdS$ solutions.

\paragraph{The G-structure of the AdS solutions} The $SU(3)$ structure
of the $AdS$ solutions, defined by all four of their Killing spinors,
is given by \cite{wrap}
\bea
J_6&=&\frac{9}{4\l}Y^aJ^a_4+\frac{9}{4\l}(1-\l^3\r^2)\frac{1}{2}\e^{abc}Y^a\mbox{D}Y^b\w\mbox{D}Y^c,\nn
\Omega_6&=&\frac{27}{8}\sqrt{\frac{1-\l^3\r^2}{\l^3}}(\e^{abc}Y^a\mbox{D}Y^b\w
J^c_4+i\mbox{D}Y^a\w J_4^a).
\eea
This structure is a solution of the $AdS$ torsion conditions of
\cite{wrap} for the near-horizon limit of fivebranes on a
co-associative four-cycle, which are
\bea
\dd\left(\frac{1}{\l^{3/2}\r}J_6\w\hat{\r}-\mbox{Im}\Omega_6\right)&=&0,\nn
\dd\left(\frac{1}{2\l}J_6\w
  J_6+\l^{1/2}\r\mbox{Re}\Omega_6\w\hat{\r}\right)&=&0.
\eea
The following identities, valid for an arbitrary conformally half-flat
Einstein manifold of scalar curvature $R$, are useful in verifying
this claim:
\bea
\dd (Y^aJ^a_4)&=&\mbox{D}Y^a\w J^a_4,\nn
\dd\left(\frac{1}{2}\e^{abc}Y^a\mbox{D}Y^b\w\mbox{D}Y^c\right)&=&\frac{R}{12}\mbox{D}Y^a\w
J^a_4,\nn
\dd (\e^{abc}Y^a\mbox{D}Y^b\w
J^c_4)&=&\frac{R}{3}\mbox{Vol}[\Sigma_4]+Y^dJ^d_4\w\e^{abc}Y^a\mbox{D}Y^b\w\mbox{D}Y^c.
\eea
In this case the Bianchi identity for the four-form is implied by the
torsion conditions \cite{wrap}.

\paragraph{The $AdS$ solution in the Minkowski frame} Using section 2,
we now
frame-rotate these solutions to the canonical Minkowski frame. The
one-form $\hat{u}$ is given by
\bea
\hat{u}=Le^{-4r/3}\dd\left(-\frac{1}{6}\sqrt{64-27\r^2}e^{-2r/3}\right).
\eea
Defining the Minkowski frame coordinate $u$,
\bea
u=-\frac{1}{6}\sqrt{64-27\r^2}e^{-2r/3},
\eea
the $AdS_3$ solutions in the Minkowski frame are given by
\bea\label{cg2ads}
\dd s^2+L^{-1}\left[\dd s^2(\mathbb{R}^{1,1})+\frac{9}{4}F\dd
  s^2(\Sigma_4)\right]+L^2\Big[F^{-4/3}\left(\dd
    u^2+u^2\mbox{D}Y^a\mbox{D}Y^a\right)+\dd s^2(\mathbb{R}^2)\Big],
\eea
where 
\bea
F&=&e^{2r},\nn
L&=&\l F,
\eea
and $e^{4r}$ is a positive signature metric inducing root of the cubic
\bea
\left(\frac{16}{9}-t^2e^{4r}\right)^3-u^6e^{4r}=0.
\eea
The wrapped-brane $G_2$ structure of the $AdS_3$ solutions is defined
by two of their Killing spinors, and is given by
\bea
\Phi&=&J_6\w\hat{u}-\mbox{Im}\Omega_6,\nn
\Upsilon&=&\frac{1}{2}J_6\w J_6+\mbox{Re}\Omega_6\w\hat{u}.
\eea
By construction, this structure is a solution of the wrapped brane
equations for fivebranes on a co-associative four-cycle. From \cite {oap}, \cite{wrap}, these
equations are
\bea
\mbox{Vol}[\mathbb{R}^2]\w\dd\Phi&=&0,\nn
\dd(L^{-1}\Phi\w\Upsilon)&=&0,\nn
\Phi\w\dd\Phi&=&0,\nn\label{cg2}
\dd\Big(L\star_9\dd(L^{-1}\Upsilon)\Big)&=&0.
\eea
In the last equation, which comes from the four-form Bianchi identity,
$\star_9$ denotes the Hodge dual on the space transverse to the
Minkowski factor.

\paragraph{The conjectured $G_2$ interpolation} We now make the
following ansatz for an interpolating solution:
\bea
\dd s^2+L^{-1}\left[\dd s^2(\mathbb{R}^{1,1})+F_1^2\dd
  s^2(\Sigma_4)\right]+L^2\left[F_3^2\dd
    u^2+F_2^2\mbox{D}Y^a\mbox{D}Y^a+\dd s^2(\mathbb{R}^2)\right],
\eea
with $L,F_{1,2,3}$ functions of $u,t$. For special holonomy 
we must have $L=\mbox{constant}$, which we take to be unity. We also
must have that $F_{1,2,3}$ are functions of $u$ only; the function
$F_3$ is then at our disposal, and we set it to $1$. The condition of
$G_2$ holonomy is then
\bea
\dd\Phi=\dd\Upsilon=0,
\eea
for the metric
\bea
\dd s^2(\mathcal{N}_{\tau})=F_1^2\dd s^2(\Sigma_4)+F_2^2\mbox{D}Y^a\mbox{D}Y^a+\dd u^2,
\eea
with the $G_2$ structure inherited from the $AdS$ frame,
\bea
\Phi&=&J_6\w\dd u-\mbox{Im}\Omega_6,\nn
\Upsilon&=&\frac {1}{2}J_6\w J_6+\mbox{Re}\Omega\w\dd u,\nn
J_6&=&F_1^2Y^aJ^a_4+\frac{1}{2}F_2^2\e^{abc}Y^a\mbox{D}Y^b\w\mbox{D}Y^c,\nn
\Omega_6&=&F_1^2F_2(\e^{abc}Y^a\mbox{D}Y^b\w
J^c_4+i\mbox{D}Y^a\w J^a_4).
\eea
With $R=-12$, closure of $\Phi$ implies
\bea\label{kl}
\partial_u(F_1^2F_2)=F_2^2-F_1^2,
\eea
while closure of $\Upsilon$ implies
\bea
\partial_u(F_1^4)&=&4F_1^2F_2,\nn\label{jk}
2\partial_u(F_1^2F_2^2)&=&-4F_1^2F_2.
\eea
It is readily verified that \eqref{jk} imply \eqref{kl}. Integrating
\eqref{jk} is straightforward. Adding, we find that
\bea
F_2^2=\frac{\a^2}{2F_1^2}-\frac{F_1^2}{2},
\eea
for some constant $\a$. Defining a new coordinate $x$ such that
\bea
\partial_u=4F_1^2F_2\partial_x,
\eea
we then get
\bea
F_1^4&=&x+\beta,
\eea
for an irrelevant constant $\beta$ which can be eliminated by a shift
in $x$. The constant $\a^2$ may be set to unity, up to an overall
scale in the metric. Defining a new coordinate $R^4=x/4$, the $G_2$ metrics conjectured
to be the interpolation of the co-associative $AdS_3$ solutions of \cite{gkw}
are as given above.

\begin{centering}
\subsection{Associative cycle}
\end{centering}
\paragraph{The interpolating pair} The $AdS_4$ solution of \cite{ach},
describing the near-horizon limit of M-fivebranes wrapped on an
associative three-cycle in a $G_2$ manifold, admits four Killing
spinors, and is as follows. The metric
is given by
\bea
\dd s^2=\frac{1}{\l}\left[\dd s^2(AdS_4)+\frac{4}{5}\dd
  s^2(H^3)+\frac{4}{25}(1-\l^3\r^2)\mu^a\mu^a+\frac{\l^3}{4}\frac{\dd
    \r^2}{1-\l^3\r^2}\right],\nonumber
\eea
\bea
\l^3=\frac{8}{5+3\r^2}.
\eea
The flux is purely magnetic and is irrelevant to the discussion; it
may be obtained from \cite{ach} or \cite{wrap}. The $\mu^a$, $a=1,2,3$, are given by
\bea
\mu^a=\s^a-\frac{1}{2}\e^{abc}\omega_{ab},
\eea
where the $\s^a$ are left-invariant one-foms on an $S^3$,
$\dd\s^a=\frac{1}{2}\e^{abc}\s^b\w\s^c$, and the $\omega_{ab}$ are the
spin-connection one-forms of $H^3$. The range of $\rho$ is
$\rho\in[-1,1]$, with the $S^3$ degenerating smoothly at
$\rho=\pm1$. \\

\noindent The conjectured $G_2$ interpolation of this metric is
\bea
\dd s^2=\dd s^2(\mathbb{R}^{1,3})+\dd s^2(\mathcal{N}_{\tau}),
\eea
where up to an overall scale
\bea\label{po}
\dd s^2(\mathcal{N}_{\tau})=\frac{R^2}{3}\dd
s^2(H^3)+\frac{R^2}{9}\Big(\frac{1}{R^3}-1\Big)\mu^a\mu^a+\left(\frac{1}{R^3}-1\right)^{-1}\dd
R^2.
\eea
This metric is singular where the associative $H^3$ degenerates, at
$R=0$. At $R=1$, the $S^3$ degenerates smoothly. This
$\mathcal{N}_{\tau}$ metric is a singular hyperbolic analogue of the
BSGPP $G_2$ metric on an $\mathbb{R}^4$ bundle over $S^3$
of \cite{bryant}, \cite{pope}. This $\mathcal{N}_{\tau}$ metric was also found in
\cite{us}, as the conjectured $G_2$ interpolation of the $AdS_2$ IIB
solution of \cite{oz}, for D3 branes wrapped on an associative
three-cycle. If it is indeed the interpolation of both these $AdS$
solutions, then there are two 
distinct conformal theories that have their origins in this
geometry. The first is a superconformal quatum mechanics, arising on
the unwrapped (time) direction of D3-branes on the $H^3$ of
\eqref{po}; the second is a three-dimensional superconformal theory,
arising on the unwrapped worldvolume directions of M5-branes 
on the $H^3$. Now we discuss the derivation of $\mathcal{N}_{\tau}$
from the M-theory $AdS_4$ solution.

\paragraph{The G-structure of the AdS solution} With $e^a$ a basis for
$H^3$, the $SU(3)$ structure of the $AdS$ solution, defined by all its
four Killing spinors, is \cite{wrap} 
\bea
J_6&=&\frac{4}{5\sqrt{5}\l}\sqrt{1-\l^3\r^2}\mu^a\w e^a,\nn
\mbox{Im}\Omega_6&=&\frac{8}{25\sqrt{5}\l^{3/2}}(1-\l^3\r^2)\frac{1}{2}\e^{abc}e^a\w\mu^b\w\mu^c-\frac{8}{5\sqrt{5}\l^{3/2}}\mbox{Vol}[H^3],\nn
\mbox{Re}\Omega_6&=&\left(\frac{2}{5\l^{1/2}}\sqrt{1-\l^3\r^2}\right)^3\frac{1}{3!}\e^{abc}\mu^a\w\mu^b\w\mu^c-\frac{8}{25}\sqrt{\frac{1-\l^3\r^2}{\l^3}}\frac{1}{2}\e^{abc}\mu^a\w
e^b\w e^c.
\eea
This structure is a solution of the $AdS$ torsion conditions of
\cite{lukas}, interpreted in \cite{wrap} as the conditions defining
the near-horizon limit of fivebranes wrapped on an associative three-cycle, which are
\bea
\dd \left(\r
  J_6\w\hat{\r}-\frac{1}{\l^{3/2}}\mbox{Im}\Omega_6\right)&=&0,\nn
\dd\left(\frac{1}{2\l\r}J_6\w
  J_6+\frac{1}{\l^{5/2}\r^2}\mbox{Re}\Omega\w\hat{\r}\right)&=&0.
\eea
Some useful identities in verifying this claim are
\bea
\dd(\mu^a\w e^a)&=&\frac{1}{2}\e^{abc}\mu^a\w\mu^b\w
e^c+3\mbox{Vol}[H^3],\nn
\dd
\left(\frac{1}{3!}\e^{abc}\mu^a\w\mu^b\w\mu^c\right)=\dd\left(\frac{1}{2}\e^{abc}\mu^a\w
  e^b\w e^c\right)&=&\frac{1}{2}e^a\w e^b\w\mu^a\w\mu^b.
\eea

\paragraph{The $AdS$ solution in the Minkowski frame} Now we use
section 2 to frame-rotate to the canonical Minkowski frame. The one-form
$\hat{u}$ is given by
\bea
\hat{u}=Le^{-3r/4}\dd u,
\eea
with the Minkowski-frame coordinate $u$ given by 
\bea
u=-\frac{4}{5}\sqrt{\frac{5-5\r^2}{8}}.
\eea
Then the associative $AdS_4$ solution in the Minkowski frame is 
\bea\label{nm}
\dd s^2=L^{-1}\left[\dd s^2(\mathbb{R}^{1,2})+\frac{4}{5}e^{2r}\dd
  s^2(H^3)\right]+L^2\left[F^{-3/4}\left(\dd
    u^2+\frac{u^2}{4}\mu^a\mu^a\right)+\dd t^2\right],
\eea
where
\bea
F&=&e^{2r},\nn
L&=&\l F,
\eea
and $e^r$ is a positive signature metric inducing root of the octic
\bea
\frac{4}{25}(1-4t^2e^{4r})^2-u^2e^{5r}=0.
\eea
The wrapped-brane $G_2$ structure of the associative $AdS_4$ solution,
defined by two of its Killing spinors, is given by
\bea
\Phi&=&J_6\w\hat{u}-\mbox{Im}\Omega_6,\nn
\Upsilon&=&\frac{1}{2}J_6\w J_6+\mbox{Re}\Omega_6\w\hat{u}.
\eea
By construction, this structure is a solution of the wrapped brane
equations for an associative three-cycle. From \cite{james},
\cite{wrap}, these are
\bea\label{ass}
\dd t\w\dd(L^{-1}\Upsilon)&=&0,\nn
\dd(L^{-5/2}\Phi\w\Upsilon)&=&0,\nn
\Phi\w\dd\Phi&=&0,\nn
\dd\Big(L^{3/2}\star_8\dd(L^{-3/2}\Phi)\Big)&=&0,
\eea
where in the last equation (the four-form Bianchi identity), $\star_8$
denotes the Hodge dual on the space transverse to the Minkowski
factor.

\paragraph{The conjectured $G_2$ interpolation} We now conjecture the
existence of a solution of \eqref{ass} which smoothly interpolates
between \eqref{nm} and a manifold of $G_2$ holonomy. We make the
following metric ansatz for this solution:
\bea 
\dd s^2=L^{-1}\left[\dd s^2(\mathbb{R}^{1,2})+F_1^2\dd
  s^2(H^3)\right]+L^2\left[F_3^2\dd
    u^2+F_2^2\mu^a\mu^a+\dd t^2\right],
\eea
with $L$, $F_{1,2,3}$ functions of $u$ and $t$. For special holonomy
we set $L=1$, and require that 
$F_{1,2,3}$ are arbitrary functions of $u$. In fact, the determination
of the $G_2$ metric from this point on exactly follows that of
\cite{us}, where a conjectured $G_2$ interpolation of the $AdS_2$
solution of \cite{oz} for a D3 brane wrapped on an associative
three-cycle was studied. The ansatz for the $G_2$ manifold is exactly
the same, and the reader is referred to \cite{us} for the rest of the
derivation, or invited to perform it as a useful excercise.

\begin{centering}
\section{Spin(7) interpolating pairs}
\end{centering}

\noindent In this section, we will give conjectured interpolating
pairs for fivebranes wrapped on Cayley four-cycles in $Spin(7)$
manifolds. First we give the pairs, then the derivation of the
$Spin(7)$ interpolations.\\

\paragraph{The interpolating pairs} The GKW $AdS_3$ solutions \cite{gkw}
describing the near-horizon limit of fivebranes on Cayley four-cycles, with membranes in the
overall transverse directions, admit two Killing spinors and have
metrics given by
\bea
\dd s^2=\frac{1}{\l}\left[\dd s^2(AdS_3)+\frac{7}{4}\dd
  s^2(\Sigma_4)+(1-\l^3f^2)\mbox{D}Y^a\mbox{D}Y^a+\frac{\l^3}{4(1-\l^3f^2)}\dd\r^2\right],\nonumber
\eea
\bea\label{spin7a}
\l^3=\frac{49}{84+15\r^2},\;\;f=\frac{6\r}{7}.
\eea     
The electric flux may be obtained from \cite{gkw} or \cite{pau}, and
the magnetic flux will be given below. The wrapped cycle $\Sigma_4$ is
an arbitrary conformally-half flat negative scalar curvature
Einstein four-manifold, normalised such that the Ricci scalar is
$R=-12$. We have flipped the definition of orientation on $\Sigma_4$
relative to \cite{gkw}; the conformally half-flat condition reads
$J^{Aij}C_{ijkl}=0$, with $J^A$, $A=1,2,3,$ a basis of self-dual
two-forms on $\Sigma_4$ and $C_{ijkl}$ the Weyl tensor on
$\Sigma_4$. The $Y^a$, $a=1,...,4$ are constrained coordinates on an
$S^3$, $Y^aY^a=1$, and
\bea
\mbox{D}Y^a=\dd
Y^a+\frac{1}{4}\omega_{cd}J^{Acd}J^{Aa}_{\;\;\;\;\;b}Y^b,
\eea
where $\omega_{ab}$ are the spin connection one-forms of
$\Sigma_4$. The range of $\rho$ is $\rho\in[-2,2]$; at the end-points,
the $S^3$ degenerates smoothly.\\

\noindent The conjectured $Spin(7)$ interpolation of this metric is
\bea
\dd s^2=\dd s^2(\mathbb{R}^{1,2})+\dd s^2(\mathcal{N}_{\tau}),
\eea
where up to an overall scale
\bea
\dd s^2(\mathcal{N}_{\tau})=\frac{9}{20}R^2\dd
s^2(\Sigma_4)+\frac{36}{100}R^2\left(\frac{1}{R^{10/3}}-1\right)\mbox{D}Y^a\mbox{D}Y^a+\left(\frac{1}{R^{10/3}}-1\right)^{-1}\dd
R^2.
\eea
These metrics are singular at $R=0$, where the Cayley four-cycle
$\Sigma_4$ degenerates. At $R=1$, the $S^3$ degenerates
smoothly. As discussed in the introduction these metrics are the analogues, for negatively curved
conformally half-flat Einstein $\Sigma_4$, of the regular BSGPP $Spin(7)$ metric
on an $\mathbb{R}^4$ bundle over $S^4$, \cite{bryant}, \cite{pope}. We
now give the derivation of the $\mathcal{N}_{\tau}$ metric from the
$AdS$ metric. 

\paragraph{The  G-structure of the AdS solution} The solution admits a $G_2$ structure,
defined by both its Killing spinors, which satisfies the torsion
conditions of \cite{james}\footnote{The conditions of \cite{james}
  contain a minor error which is corrected in \cite{pau}.} together
with the Bianchi identity for the magnetic flux (also given in
\cite{james}) which in this case is not implied by the torsion
conditions. the torsion conditions of \cite{james} were interpreted in
\cite{pau} as the conditions defining the near-horizon limit of
fivebranes wrapped on a Cayley four-cycle. These conditions are satisfied by all supersymmetric
$AdS_3$ solutions of M-theory. If $e^a$ denote a basis for $\Sigma_4$, the $G_2$
structure of the $AdS$ solutions is given by
\bea
\Phi&=&-\frac{7}{4}\sqrt{\frac{1-\l^3f^2}{\l^3}}\left[Y^ae^a\w
  e^b\w\mbox{D}Y^b+\frac{1}{2}\e^{abcd}Y^a\mbox{D}Y^b\w e^c\w
  e^d\right]\nn&&\label{structure}+\left(\sqrt{\frac{1-\l^3f^2}{\l}}\right)^3\frac{1}{3!}\e^{abcd}Y^a\mbox{D}Y^b\w\mbox{D}Y^c\w\mbox{D}Y^d,\\
\Upsilon&=&-\frac{7}{4\l^2}(1-\l^3f^2)\left[\frac{1}{2}e^a\w
  e^b\w\mbox{D}Y^a\w\mbox{D}Y^b+\frac{1}{4}\e^{abcd}\mbox{D}Y^a\w\mbox{D}Y^c\w
  e^c\w e^d\right]+\frac{49}{16\l^2}\mbox{Vol}[\Sigma_4].\nn
\eea  
The torsion conditions of \cite{james} are
\bea
\hat{\r}\w\dd(\l^{-1}\Upsilon)&=&0,\\
\dd\left(\l^{-5/2}\sqrt{1-\l^3f^2}\mbox{Vol}[\mathcal{M}_7]\right)&=&-4\l^{-1/2}f\hat{\r}\w\mbox{Vol}[\mathcal{M}_7],\\
\dd\Phi\w\Phi&=&\frac{4\l^{1/2}}{\sqrt{1-\l^3f^2}}(3-\l^3f^2)\mbox{Vol}[\mathcal{M}_7]-2\l^{3/2}f\star_8\dd\log\left(\frac{\l^3f}{1-\l^3f^2}\right),\nn
\eea
where
\bea
\mbox{Vol}[\mathcal{M}_7]=\frac{1}{7}\Phi\w\Upsilon.
\eea
The four-form Bianchi identity is $\dd F_{\mbox{\small{mag}}}=0$, with
\bea
F_{\mbox{\small{mag}}}&=&
\frac{\l^{3/2}}{\sqrt{1-\l^3f^2}}\Big(\l^{3/2}f+\star_8\Big)\Big(\dd[\l^{-3/2}\sqrt{1-\l^3f^2}\Phi]-4\l^{-1}\Upsilon\Big)+2\l^{1/2}\Phi\w\hat{\r},
\eea
where $\star_8$ denotes the Hodge dual on the space transverse to the
$AdS_3$ factor, with positive orientation defined with respect to
\bea
\mbox{Vol}=\mbox{Vol}[\mathcal{M}_7]\w\hat{\r}.
\eea 
It may be verified that the structure
\eqref{structure} is indeed a solution of the torsion conditions and
Bianchi identity, by using
the following identities, valid for any conformally half-flat Einstein
$\Sigma_4$ with scalar curvature $R$:
\bea
&&\dd\left[Y^ae^a\w
  e^b\w\mbox{D}Y^b+\frac{1}{2}\e^{abcd}Y^a\mbox{D}Y^b\w e^c\w
  e^d\right]\nn&&=-\frac{R}{4}\mbox{Vol}[\Sigma_4]+e^a\w
  e^b\w\mbox{D}Y^a\w\mbox{D}Y^b+\frac{1}{2}\e^{abcd}\mbox{D}Y^a\w\mbox{D}Y^c\w
  e^c\w e^d,\nn&&
\dd\left[\frac{1}{3!}\e^{abcd}Y^a\mbox{D}Y^b\w\mbox{D}Y^c\w\mbox{D}Y^d\right]\nn&&\label{poiu}=-\frac{R}{48}\left[e^a\w
  e^b\w\mbox{D}Y^a\w\mbox{D}Y^b+\frac{1}{2}\e^{abcd}\mbox{D}Y^a\w\mbox{D}Y^c\w
  e^c\w e^d\right].
\eea

\paragraph{The AdS solutions in the Minkowski frame} Defining the coordinates
\bea
t&=&-\frac{1}{2}e^{-12r/7}\r,\nn
u&=&-\sqrt{\frac{12-3\r^2}{7}}e^{-r},
\eea
the one-forms $e^8$, $e^9$ in the Minkowski frame are given by
\bea
e^8&=&\l e^r\dd u,\nn
e^9&=&\l e^{12r/7}\dd t,
\eea
and the metric in the Minkowski frame takes the form
\bea
\dd s^2&=&\frac{1}{H^{1/3}_{M5}H_{M2}^{2/3}}\dd
s^2(\mathbb{R}^{1,1})+\frac{H^{2/3}_{M5}}{H^{2/3}_{M2}}\dd
t^2+\frac{H^{1/3}_{M2}}{H^{1/3}_{M5}}\left[\frac{7}{4}F\dd
  s^2(\Sigma_4)\right]\nn&&+H^{1/3}_{M2}H_{M5}^{2/3}\left[\frac{1}{F}\left(\dd
    u^2+u^2\mbox{D}Y^a\mbox{D}Y^a\right)\right],
\eea
where
\bea
H_{M5}&=&\l^3 e^{38r/7},\nn
H_{M2}&=&e^{2r/7},\nn
F&=&e^{12r/7}.
\eea 
The function $e^{2r}$ is given in terms of $t$ and $u$ by a positive
signature metric inducing root of the twelfth order polynomial
\bea
t^{14}e^{24r}-\left(1-\frac{7}{12}u^2e^{2r}\right)^7=0.
\eea
The wrapped brane $Spin(7)$ structure of the $AdS_3$ solutions,
defined by one of their Killing spinors, is
given by
\bea
\phi=-\Phi\w e^8-\Upsilon.
\eea
By construction, this structure is a solution of the wrapped brane
equations for a Cayley four-cycle in a $Spin(7)$ manifold, which
comprise the torsion conditions \cite{j}, \cite{pau}
\bea
e^9\w\left[-L^3e^9\lrcorner\; \dd
  (L^{-3}e^9)+\frac{1}{2}\phi\lrcorner\;\dd\phi\right] &=&0,\\
(e^9\w+\star_9)[e^9\lrcorner\;\dd (L^{-1}\phi)]&=&0,
\eea  
together with the Bianchi identity and field equation for the
four-form, which is given in \cite{j}, \cite{pau}.

\paragraph{The conjectured Spin(7) interpolation} We make the
following ansatz for an interpolating solution:
\bea
\dd s^2&=&\frac{1}{H^{1/3}_{M5}H_{M2}^{2/3}}\dd
s^2(\mathbb{R}^{1,1})+\frac{H^{2/3}_{M5}}{H^{2/3}_{M2}}\dd
t^2+\frac{H^{1/3}_{M2}}{H^{1/3}_{M5}}\left[F_1^2\dd
  s^2(\Sigma_4)\right]\nn&&+H^{1/3}_{M2}H_{M5}^{2/3}\left[F_3^2\dd
    u^2+F_2^2\mbox{D}Y^a\mbox{D}Y^a\right],
\eea   
with $H_{M5,M2}$, $F_{1,2,3}$ arbitrary functions of $u,t$. To
determine the $Spin(7)$ interpolation with this ansatz,
we set $H_{M5,M2}=1$ and require that $F_{1,2,3}$ are functions only of
$u$. Then $F_3$ is at our disposal, and we set it to 1. Requiring
$Spin(7)$ holonomy, we set
\bea
\dd\phi=0,
\eea
with 
\bea
\phi&=&-\Phi\w\dd u-\Upsilon,\nn
\Phi&=&-F_1^2F_2\left[Y^ae^a\w
  e^b\w\mbox{D}Y^b+\frac{1}{2}\e^{abcd}Y^a\mbox{D}Y^b\w e^c\w
  e^d\right]+F_2^3\frac{1}{3!}\e^{abcd}Y^a\mbox{D}Y^b\w\mbox{D}Y^c\w\mbox{D}Y^d,\nn
\Upsilon&=&-F_1^2F_2^2\left[\frac{1}{2}e^a\w
  e^b\w\mbox{D}Y^a\w\mbox{D}Y^b+\frac{1}{4}\e^{abcd}\mbox{D}Y^a\w\mbox{D}Y^c\w
  e^c\w e^d\right]+F_1^4\mbox{Vol}[\Sigma_4].
\eea  
 Using \eqref{poiu} with $R=-12$, the $Spin(7)$ condition reduces to
\bea
\partial_u(F_1^4)&=&3F_1^2F_2,\nn
\frac{1}{2}\partial_u(F_1^2F_2^2)&=&\frac{1}{3}F_2^3-F_1^2F_2.
\eea
Defining a new coordinate $x$ such that
\bea
\partial_u=\frac{3}{4}\partial_x,
\eea
we get
\bea 
F_1=x+\a,
\eea
for a constant $\a$ which may be eliminated by a shift in $x$. Then
\bea
F^2_2=\frac{1}{x^{4/3}}\left(\beta-\frac{4}{5}x^{10/3}\right),
\eea
for a constant $\beta$ which may be set to unity up to an overall scale
in the metric. Defining a new coordinate $x^{10/3}=5R^{10/3}/4$, up
to an overall scale we obtain the $\mathcal{N}_{\tau}$ metric given above.

\begin{centering}
\section{Conclusions and outlook}
\end{centering}

\noindent In this paper, the notion of an interpolation between
Anti-de Sitter
and special holonomy manifolds has been defined. The importance of
this concept in the geometry of the supersymmetric AdS/CFT
correspondence has been stressed. Two conjectures have been made: that
all supersymmetric $AdS$ solutions of M-/string theory admit a special
holonomy interpolation, and that, with the exception of flat space,
all metrics on special holonomy manifolds admitting an $AdS$ interpolation are
incomplete. For a representative sample of known
supersymmetric $AdS$ solutions of M-theory, a series of canditate
incomplete special holonomy interpolations has been derived. The
series of interpolations is closely related to a set of celebrated
complete special holonomy metrics.

Several interesting directions for future research are suggested by
the results of this paper. The geometrical question of most importance
is undoubtedly the construction of an interpolating solution describing a
wrapped brane, for one of the proposed interpolating pairs of this
paper. Since the whole series of pairs share many common features,
understanding how to do this for one of them would almost certainly
facilitate the construction of an interpolating solution for all. A
reasonable guess for what the boundary conditions of an interpolating
solution for these pairs should be is the following. It should match
on to an $\mathcal{N}_{\tau}$ metric at its regular degeneration
point. It should also match on the $AdS$ solution at a degeneration
point of its transverse space. There is an unfixed volume modulus in
all of the $\mathcal{N}_{\tau}$ metrics; this will be fixed, in an
interpolating solution, by the global topological requirement of
matching onto an $AdS$ solution. For the $AdS$ solutions without
R-symmetry isometries, the degeneration points of the transverse space
are symmetric; an interpolating solution should match on to one of
them. For the $AdS$ solutions with R-symmetry isometries, the
degeneration points of the transverse space are asymmetric; in this
case, it seems plausible that an interpolating solution should match
on to the $AdS$ solution at its R-symmetry degeneration
point. Understanding how this comes about, and solving the wrapped
brane equations for an interpolating solution, is not just a problem
in Riemannian geometry. It seems very likely that the {\it Lorenztian}
character of an interpolating solution will enter in an essential way,
with the causal structure of the interpolating solution playing a key
part. This is because (at least by analogy with conical
interpolations) an interpolating solution should match on to the
special holonomy manifold at a spacelike infinity, and the $AdS$
manifold at an event horizon. Of the two coordinates which play a
r$\hat{\mbox{o}}$le 
in the frame rotation underlying the relationship between the
interpolating pairs of this paper, one has a finite range while the
range of the other is infinite. Though they cannot really be
seperated, in a rough sense the non-compact direction should determine
the Lorentzian, causal structure, and the compact direction the
Riemannian. A very delicate interplay between the
two is required, to fulfill the appropriate Lorenztian and Riemannian
boundary conditions for an interpolating solution. Understanding the
geometry of the frame rotation in more depth may reveal how to
linearise the wrapped brane equations, and so superimpose the
interpolating pair, just as for conical interpolations. Another
intriguing point about the frame rotation is that the relationship
between the $AdS$ and Minkowski frame coordinates is in every case
given by the root of a polynomial. This strongly suggests some deeper
underlying algebraic geometry which has not been appreciated.

Other interesting geometrical questions raised by this work include
the following. For branes wrapped on K\"{a}hler cycles, there exist
rich classes of $AdS$ solutions that have not been studied here. These
include $AdS_5$ solutions from M-fivebranes on two-cycles in
three-folds \cite{ypq}, $AdS_3$ solutions from M-fivebranes on
four-cycles in fourfolds \cite{toni1}, \cite{dan} and $AdS_3$ solutions from
D3-branes on two-cycles in four-folds \cite{toni2}, \cite{dan}. It would be
interesting to apply the methods of this paper to these other solutions,
and so determine candidate interpolations. For the $AdS$-from-D3-brane
solutions of \cite{toni2}, \cite{dan} it should be particularly feasible to
construct the interpolating solutions, since in this case the
four-fold geometry is essentially conical \cite{nakwoo}, \cite{dan},
\cite{us}. Also $AdS_2$ M-theory solutions have not been discussed in
this paper at all; a rich class has recently been discovered in
\cite{dan}, and some older ones are to be found in \cite{gkw}. Using
the classification results of \cite {nakwoo1}, \cite{eoin}, it would
be interesting to determine their candidate interpolations.

It should also be possible to use the notion of an interpolating pair
to construct new $AdS$ solutions. For all cases other than K\"{a}hler
cycles, to the knowledge of the author, only a single $AdS$ solution
is known to exist - the one studied in this paper. On the other hand,
numerous complete cohomogeneity-one special holonomy metrics are known; for example, for
$G_2$ and $Spin(7)$, several complete metrics, whose construction was
inspired by the BSGPP metrics, were given in \cite{cvetic2},
\cite{gukov}. Hyperbolic analogues of these metrics should also exist,
and if so, it will almost certainly be possible to map them to new
$AdS$ solutions.  

A more long-term project concerns the construction of the conformal
quantum duals of the interpolating pairs. In M-theory, this problem is
hampered by the notoriously intractable question of the effective
field theory on the worldvolume of a stack of fivebranes (for
membranes, some interesting progress on the world-volume theory,
highlighting its non-associativity, has recently been made in
\cite{lambert}). In IIB, this is less of a problem, and it should be
possible to make progress constructing the duals of wrapped D3-brane
geometries, even with existing techniques. 

In the geometry of wrapped brane physics, we have for so long been
restricted to the near-horizon limit, the $AdS$ geometry, that it has
become commonplace to think that {\it only} this geometry is of
relevance to investigations of the CFT. Indeed, recently it has been
shown that it is in fact possible in principle to reconstruct the CFT
from the near-horizon geometry alone\footnote{I thank Marika Taylor
  for pointing this out to me.} using holographic renormalisation
techniques \cite{skenderis}, \cite{marika}. However, doing
this for $AdS$ geometries of the complexity of those studied in this
paper is likely to be very difficult indeed, if not impossible, in
practice. And focussing on the $AdS$ geometry alone ignores the
central message of this paper: that the geometry of AdS/CFT involves,
in an essential way, {\it both} an Anti-de Sitter and a special
holonomy manifold. It is also possible, as a matter of principle, to
construct the CFT dual from the geometry of the special holonomy
manifold alone. It is worth recalling that this is how the
quiver gauge theory duals of the $Y^{p,q}$ manifolds were in fact constructed
\cite{3}, \cite{4}; as, indeed, was $\mathcal{N}=4$ super Yang Mills
itself in this context \cite{malda}. Knowing both members of an
interpolating pair means that CFT construction techniques can be
brought to bear on both geometries; knowing both significantly
enriches our understanding of the correspondence.

\begin{centering}
\section{Acknowledgements}
\end{centering}
\noindent I would like to express my gratitude to Jerome Gauntlett for
ongoing discussions, collaboration and debate; and also to Dina
Daskalopoulou, for general inspiration. This work was supported by EPSRC.


\begin{thebibliography}{99}

\bibitem{malda} J. M. Maldacena, ``The Large N Limit of Superconformal
  Field Theories and Supergravity'', Adv.Theor.Math.Phys. {\bf 2} (1998)
  231-252; Int.J.Theor.Phys. {\bf 38} (1999) 1113-1133, hep-th/9711200.

\bibitem{beisert} N. Beisert, ``The S-Matrix of
  AdS/CFT and Yangian symmetry'', PoS (Solvay) {\bf 002} (2007),
  arXiv:0705.0321.

\bibitem{zarembo} K. Zarembo, ``Semiclassical Bethe ansatz and
  AdS/CFT'', Comptes Rendus Physique {\bf 5} (2004) 1081,
  Fortsch.Phys. {\bf 53} (2005) 647.

\bibitem{26} J. Maldacena and C. Nu\~nez, ``Towards the large N limit
  of pure N=1 super Yang Mills'', Phys.Rev.Lett. {\bf 86} (2001) 588-591,
  hep-th/0008001. 

\bibitem{kiritsis} U. Gursoy and E. Kiritsis, ``Exploring improved
  holographic theories for QCD: Part 1'', arXiv:0707.1324.

\bibitem{k1} U. Gursoy,
  E. Kiritsis and F. Nitti, ``Exploring improved holographic theories
  for QCD: part II'', arXiv:0707.1349.

\bibitem{sean} S. A. Hartnoll and C. P. Herzog, ``Ohm's law at strong
  coupling: S duality and the cyclotron resonance'', arXiv:0706.3228.

\bibitem{uiy}  S. A. Hartnoll, P. K. Kovtun, M. Mueller and S. Sachdev, ``Theory of
  the Nernst effect near quantum phase transitions in condensed
  matter, and in dyonic black holes'', arXiv:0706.3215.
 
\bibitem{lks} S. A. Hartnoll
  and P. Kovtun, ``Hall conductivity from dyonic black holes'',
  arXiv:0704.1160. 

\bibitem{rhic} D. Mateos, R. C. Myers and R. M. Thomson, ``Holographic
  viscosity of fundamental matter'', Phys. Rev. Lett {\bf 98} (2007),
  101601, hep-th/0610184.

\bibitem{ks} I. R. Klebanov and M. J. Strassler, ``Supergravity and a
  Confining Gauge Theory: Duality Cascades and $\chi$SB-Resolution of
  Naked Singularities'', JHEP {\bf 0008} (2000) 052, hep-th/0007191.

\bibitem{kleb} I. R. Klebanov and E. Witten, ``Superconformal Field
  Theory on Threebranes at a Calabi-Yau Singularity'', Nucl.Phys. B {\bf
    536}
  (1998) 199-218, hep-th/9807080.

\bibitem{ypq} J. P. Gauntlett, D. Martelli, J. Sparks and D. Waldram,
  ``Supersymmetric AdS5 solutions of M-theory'',
  Class.Quant.Grav. {\bf 21}
  (2004) 4335-4366, hep-th/0402153.

\bibitem{sas} J. P. Gauntlett, D. Martelli, J. Sparks and D. Waldram,
  ``Sasaki-Einstein Metrics on $S^2 x S^3$'',
  Adv.Theor.Math.Phys. {\bf 8}
  (2004) 711-734, hep-th/0403002.

\bibitem{2} B. Feng, A. Hanany and Y.-H. He, ``D-Brane Gauge Theories
  from Toric Singularities and Toric Duality'', Nucl.Phys. B {\bf 595} (2001)
  165-200, hep-th/0003085.


\bibitem{3} D. Martelli and J. Sparks, ``Toric Geometry,
  Sasaki-Einstein Manifolds and a New Infinite Class of AdS/CFT
  Duals'', Commun.Math.Phys. {\bf 262} (2006) 51-89, hep-th/0411238. 

\bibitem{4} S. Benvenuti, S. Franco, A. Hanany, D. Martelli and
  J. Sparks, ``An Infinite Family of Superconformal Quiver Gauge
  Theories with Sasaki-Einstein Duals'', JHEP {\bf 0506} (2005) 064,
  hep-th/0411264. 

\bibitem{5} K. Intrilligator and B. Wecht, ``The Exact Superconformal
  R-Symmetry Maximizes a'', Nucl.Phys. B{\bf 667} (2003) 183-200, hep-th/0304128.

\bibitem{6} D. Martelli, J. Sparks and S.-T. Yau, ``The Geometric Dual
  of a-maximisation for Toric Sasaki-Einstein Manifolds'',
  Commun.Math.Phys. {\bf 268} (2006) 39-65, hep-th/0503183.

\bibitem{gibbons} M. J. Duff, G. W. Gibbons and P. K. Townsend, ``Macroscopic
superstrings as interpolating solitons'', Phys. Lett. B {\bf  332}
(1994) 321-328, hep-th/9405124.

\bibitem{mn} J. Maldacena and C. Nu\~nez, ``Supergravity description
  of field theories on curved manifolds and a no go theorem'',
  Int.J.Mod.Phys.  {\bf A 16} (2001) 822-855, hep-th/0007018.

\bibitem{ach} B. S. Acharya, J. P. Gauntlett and N. Kim, ``Fivebranes
  wrapped on associative three-cycles'', Phys. Rev. D {\bf 63} (2001)
  106003, hep-th/0011190.

\bibitem{sezgin} M. Pernici and E. Sezgin, ``Spontaneous
  compactification of seven-dimensional supergravity theories'',
  Class. Quant. Grav. {\bf 2} (1985), 673.

\bibitem{gkw} J. P. Gauntlett, N. Kim and D. Waldram, ``M-fivebranes
  wrapped on supersymmetric cycles'', Phys. Rev. D {\bf 63} (2001)
  126001, hep-th/0012195.

\bibitem{jk} J. P. Gauntlett and N. Kim, ``M-fivebranes wrapped on
  supersymmetric cycles II'', Phys. Rev.  D{\bf 65} (2002) 086003,
  hep-th/0109039.  

\bibitem{oz} H. Nieder and Y. Oz, ``Supergravity and D-branes wrapping
  supersymmetric cycles'', JHEP {\bf 0103} (2001) 008, hep-th/0011288.

\bibitem{naka}
  M. Naka,
  ``Various wrapped branes from gauged supergravities,''
  hep-th/0206141.

\bibitem{toni1} J. P. Gauntlett, O. A. P. Mac Conamhna, T. Mateos and
  D. Waldram, ``New supersymmetric AdS3 solutions'', Phys.Rev. D {\bf 74}
  (2006) 106007, hep-th/0608055.

\bibitem{toni2} J. P. Gauntlett, O. A. P. Mac Conamhna, T. Mateos and
  D. Waldram, ``Supersymmetric AdS3 solutions of type IIB
  supergravity'', Phys.Rev.Lett {\bf 97} (2006) 171601,
  hep-th/0606221.

\bibitem{dan}
  J. P. Gauntlett, N. Kim and D. Waldram, ``Supersymmetric AdS(3),
  AdS(2) and bubble solutions,'' JHEP {\bf 0704} (2007) 005,
  hep-th/0612253. 

\bibitem{me} O. A. P. Mac Conamhna, ``Inverting geometric transitions:
  explicit Calabi-Yau metrics for the Maldacena-Nu\~{n}ez solutions'',
  arXiv:0706.1795.

\bibitem{smith} A. Fayyazuddin and D. J. Smith, 
   ``Localized intersections of M5-branes and four-dimensional
   superconformal field theories'',  
  JHEP {\bf 9904} (1999) 030, [hep-th/9902210].

\bibitem{kastor} H. Cho, M. Emam, D. Kastor and J. Traschen,
  ``Calibrations and Fayyazuddin-Smith Spacetimes'', Phys.Rev. D {\bf 63}
  (2001) 064003, hep-th/0009062.

\bibitem{Husain:2003df}
  T.~Z.~Husain,
  ``That's a wrap!'',
  JHEP {\bf 0304} (2003) 053, hep-th/0302071.
  %%CITATION = HEP-TH 0302071;%%


\bibitem{smith1} B. Brinne, A. Fayyazuddin, T. Z. Husain and
  D. J. Smith, ``$N=1$ M5-brane geometries'', JHEP {\bf 0103} (2001) 052,
 hep-th/0012194.


\bibitem{fayy3} A. Fayyazuddin, T. Z. Husain and I. Pappa, ``The
  geometry of wrapped M5-branes in Calabi-Yau 2-folds'', hep-th/0509018.

\bibitem{james} D. Martelli and J. Sparks, ``G-structures, fluxes and
  calibrations in M-theory'', Phys. Rev. D {\bf 68} (2003) 085014,
  hep-th/0306225.

\bibitem{43} T. Z. Husain, ``M2-branes wrapped on holomorphic
  curves'', JHEP {\bf 0312} (2003) 037, hep-th/0211030.
  
\bibitem{wrap}
  J. P. Gauntlett, O. A. P. Mac Conamhna, T. Mateos and D. Waldram,
  ``AdS spacetimes from wrapped M5 branes,''
  JHEP {\bf 0611} (2006) 053, hep-th/0605146.
 
\bibitem{eoin} O. A. P. Mac Conamhna and E. \'{O} Colg\'{a}in,
  ``Supersymmetric wrapped membranes, AdS(2) spaces, and bubbling
  geometries'', JHEP {\bf 0703} (2007) 115, hep-th/0612196.

\bibitem{pau} P. Figueras, O. A. P. Mac Conamhna and E. \'{O}
  Colg\'{a}in, ``Global geometry of the supersymmetric $AdS_3/CFT_2$
  correspondence in M-theory'', Phys. Rev. D {\bf 76} (2007) 046007,
  hep-th/0703275.

\bibitem{eguchi} T. Eguchi and A. J. Hanson, ``Asymptotically flat
  self-dual solutions to Euclidean gravity'', Phys. Lett. B {\bf 74},
  249 (1978).

\bibitem{bryant} R. L. Bryant and S. Salamon, ``On the construction of
  some complete metrics with exceptional holonomy'', Duke
  Math. J. {\bf 58}, 829 (1989).

\bibitem{pope} G. W. Gibbons, D. N. Page and C. N. Pope, ``Einstein
  metrics on $S^3$, $R^3$ and $R^4$ bundles'',
  Commun. Math. Phys. {\bf 127}, 529 (1990).

\bibitem{cand}
  P. Candelas and X. C. de la Ossa,
  ``Comments on Conifolds,'' Nucl. Phys. B {\bf 342} (1990) 246.

\bibitem{pando}
  L. A. Pando Zayas and A. A. Tseytlin,
  ``3-branes on resolved conifold,''
  JHEP {\bf 0011} (2000) 028, hep-th/0010088.

\bibitem{tseytlin} G. Papadopoulos and A. A. Tseytlin, ``Complex
  geometry of conifolds and 5-brane wrapped on 2-sphere'',
  Class. Quant. Grav. {\bf 18} (2001) 1333, hep-th/0012034.

\bibitem{stenzel} M. B. Stenzel, ``Ricci-flat metrics on the
  complexification of a compact rank one symmetric space'',
  Manuscripta Mathematica {\bf 80}, 151 (1993).

\bibitem{cvetic} M. Cvetic, G. W. Gibbons, H. L\"{u} and C. N. Pope,
  ``Ricci-flat metrics, harmonic forms and brane resolutions'',
  Commun.Math.Phys. {\bf 232} (2003) 457, hep-th/0012011. 

\bibitem{cvetic1}  M. Cvetic, G. W. Gibbons, H. L\"{u} and C. N. Pope,
  ``Special holonomy spaces and M-theory'',  hep-th/0206154.

\bibitem{ohta} K. Ohta and T. Yokono, ``Deformation of conifold and
  intersecting branes'', JHEP {\bf 0002} (2000) 023, hep-th/9912266.

\bibitem{cvetic2} M. Cvetic, G. W. Gibbons, H. L\"{u} and C. N. Pope,
  ``Cohomogeneity one manifolds of Spin(7) and $G_2$ holonomy'',
  Phys. Rev. D{\bf 65} (2002) 106004, hep-th/0108245. 

\bibitem{gukov} S. Gukov and J. Sparks, ``M-theory on Spin(7)
  manifolds'', Nucl. Phys. B {\bf 625} (2002) 3, hep-th/0109025. 

\bibitem{calabi} E. Calabi, ``M\'{e}triques K\"{a}hl\'{e}riennes et
  fibr\'{e}s holomorphe'', Ann. Scient. \'{E}cole Norm. Sup., {\bf
    12}, 269 (1979).

\bibitem{swann} A. Dancer and A. Swann, ``Hyperk\"{a}hler metrics of
  cohomogeneity one'', J. Geometry and Physics {\bf 21}, 218 (1997).

\bibitem{cvetic3}  M. Cvetic, G. W. Gibbons, H. L\"{u} and C. N. Pope,
  ``Hyper-K\"{a}hler Calabi Metrics, $L^2$ harmonic forms, resolved
  M2-branes, and $AdS_4/CFT_3$ correspondence'', Nucl. Phys. B {\bf
    617} (2001) 151, hep-th/0102185.

\bibitem{oap} O. A. P. Mac Conamhna, ``The geometry of extended null
  supersymmetry in M-theory'', Phys. Rev. D {\bf 73} (2006) 045012,
  hep-th/0505230. 


\bibitem{lukas} A. Lukas and P. Saffin, ``M-theory compactification,
  fluxes and $AdS_4$'', Phys. Rev. D {\bf  71} (2005) 046005, hep-th/0403235.


\bibitem{us} J. P. Gauntlett and O. A. P. Mac Conamhna, ``AdS
  spacetimes from wrapped D3 branes'', arXiv:hep-th/0707.3105.

\bibitem{oap2} O. A. P. Mac Conamhna, ``Eight-manifolds with
  G-structure in eleven dimensional supergravity'', Phys.Rev. D {\bf 72}
  (2005) 086007, hep-th/0504028.


\bibitem{j} J. P. Gauntlett, J. B. Gutowski and S. Pakis, ``The
  Geometry of D=11 Null Killing Spinors'', JHEP {\bf 0312} (2003) 049,
  hep-th/0311112.

\bibitem{nakwoo}
  N. Kim, ``AdS(3) solutions of IIB supergravity from D3-branes,''
  JHEP {\bf 0601} (2006) 094, hep-th/0511029.

\bibitem{nakwoo1}
  N. Kim and J. D. Park, ``Comments on AdS(2) solutions of D = 11
  supergravity,'' JHEP {\bf 0609} (2006) 041, hep-th/0607093.

\bibitem{lambert} J. Bagger and N. Lambert, ``Modelling multiple
  M2's'', Phys. Rev. D {\bf 75}, (2007) 045020, hep-th/0611108.

\bibitem{skenderis} K. Skenderis, ``Lecture notes on holographic
  renormalisation'', Class. Quant. Grav. {\bf 19} (2002) 5849,
  hep-th/0209067.  

\bibitem{marika} K. Skenderis and M. Taylor, ``Kaluza-Klein
  holography'', JHEP {\bf 0605} (2006) 057, hep-th/0603016.

\end{thebibliography}
\end{document}